%% file: manuscript_arxiv.tex
\pdfoutput=1
\documentclass[%
 reprint,
showpacs,
 amsmath,amssymb,
 aps,
pra,
floatfix,
]{revtex4-1}

\usepackage{graphicx,epstopdf}
\graphicspath{ {figures/} } 
\usepackage{dcolumn}
\usepackage{bm}

\usepackage{mathtools}\usepackage{float}
\DeclareGraphicsExtensions{.ps}
\usepackage{verbatim}		
\usepackage[caption=false]{subfig}
\usepackage{amsmath}
\usepackage{amssymb}
\usepackage{amsthm}			
\usepackage[english]{babel}

\usepackage[colorlinks=true, linkcolor=blue, citecolor=blue, urlcolor=blue, filecolor=blue]{hyperref}	

\begin{document}

\preprint{APS/123-QED}

\title{Scale-Free Crystallization of two-dimensional Complex Plasmas: Domain Analysis using Minkowski Tensors}

\author{A. B\"obel}
	\email{alexander.boebel@dlr.de}
\author{C. A. Knapek}%
\author{C. R\"ath}%
 
\affiliation{%
 Institut f\"ur Materialphysik im Weltraum, Deutsches Zentrum f\"ur Luft- und Raumfahrt (DLR), M\"unchner Str. 20, 82234 We{\ss}ling\\
}%




\date{\today}


\renewcommand{\figurename}{FIG.}

\begin{abstract}
Experiments of the recrystallization processes in two-dimensional complex plasmas are analyzed in order to rigorously test a recently developed scale-free phase transition theory. The "Fractal-Domain-Structure" (FDS) theory is based on the kinetic theory of Frenkel. It assumes the formation of homogeneous domains, separated by defect lines, during crystallization and a fractal relationship between domain area and boundary length. For the defect number fraction and system energy a scale free power-law relation is predicted.

The long range scaling behavior of the bond order correlation function shows clearly that the complex plasma phase transitions are not of KTHNY type. Previous preliminary results obtained by counting the number of dislocations and applying a bond order metric for structural analysis are reproduced. These findings are supplemented by extending the use of the bond order metric to measure the defect number fraction and furthermore applying state-of-the-art analysis methods, allowing a systematic testing of the FDS theory with unprecedented scrutiny: A morphological analysis of lattice structure is performed via Minkowski tensor methods. Minkowski tensors form a complete family of additive, motion covariant and continuous morphological measures that are sensitive to non-linear properties. The FDS theory is rigorously confirmed and predictions of the theory are reproduced extremely well. The predicted scale-free power law relation between defect fraction number and system energy is verified for one more order of magnitude at high energies compared to the inherently discontinuous bond order metric.

It is found that the the fractal relation between crystalline domain area and circumference is independent of the experiment, the particular Minkowski tensor method and the particular choice of parameters. Thus, the fractal relationship seems to be inherent to two-dimensional phase transitions in complex plasmas.

Minkowski Tensor analysis turns out to be a powerful tool for investigations of crystallization processes. It is capable to reveal non-linear local topological properties, however, still provides easily interpretable results founded on a solid mathematical framework.

\end{abstract}

\maketitle


\section{Introduction \label{sec1}}

Complex plasmas are composed of a weakly ionized gas and micro-particles which are highly charged due to absorption of the ambient electron- and ion-streams \cite{dmx10, 12o}. Complex plasmas constitute a model system which is well suited for studying the kinetics of fluids and crystallization processes at the individual particle level in three or two dimensions. Properties of pair interactions, such as the interaction range and strength, can be flexibly tuned. Also, the dynamics of particles at short time scales is practically undamped due to the low gas density in typical complex plasmas \cite{12o}.

Because the Mermin-Wagner \cite{01o} theorem forbids any long-range order in only two dimensions the existence of crystallization in two-dimensional phase transitions seemed thermodynamically impossible. However, Kosterlitz and Thouless proved \cite{02o,03o,04o,05o,06o} the possibility of a topological phase transition, from solid to liquid, in two-dimensional systems. This KT transition is mediated by lattice defects. Paired dislocations as initially bound defects dissociate into an intermediate hexatic phase that consists mainly of free dislocations, which then dissociate into free disclinations as the liquid state is reached. A disclination is a crystal defect for which rotational symmetry is broken. A dislocation is a type of defect that breaks translational symmetry. The general term defect refers to either or a combination of both types. In the KT transition the long-range order typical to three-dimensional crystals is replaced by a quasi-long-range order. Thus the Mermin-Wagner-Theorem \cite{01o} is not violated. Experimental evidence for such a topological phase transition is rare, examples are e.g. colloidal systems \cite{07o,08o}, the two-dimensional electron sheet on liquid helium \cite{09o}, atomic gases \cite{10o} and superconducting vortex lattices \cite{11o}. However, only recently it was shown for a colloidal suspension system, that the conventional KTHNY theory is not applicable on spherical geometry \cite{Guerra2018}.

Whether a phase transition is of KTHNY type can be reduced to the question whether the pair correlation function $g(r)$ or the bond correlation function $g_6(r)$ follows a specific scaling behavior \cite{kthny_strandburg, kthny_alba}.
In \cite{15o} it is shown, by both experimental and simulated data, that the recrystallization of two-dimensional complex plasmas is not compatible with the KTHNY theory of phase transition due to different scaling behaviors in $g(r)$ and $g_6(r)$. In this work a "Fractal-Domain-Structure" (FDS) theory \cite{12o} based on the kinetic theory of Frenkel \cite{13o} is tested. It assumes the formation of homogeneous domains, separated by defect lines, during crystallization. Based on experimental evidence, a fractal relationship between domain area and boundary length is postulated. For the defect number fraction and system energy a scale free power-law relation is predicted.

The FDS theory is tested for experiments and a simulation of the crystallization process in two-dimensional complex plasmas. A layer of micro-particles is levitated in the plasma sheath region and illuminated by a thin laser sheet. The crystalline particle system is melted by a short electric pulse and the recrystallization is captured by a high speed camera.

Indications that this complex plasma phase transition data confirm the FDS theory were given in a first study \cite{knapek_rec}. There, defect numbers were counted as 5/7-dislocations. The hexagonal translational order in the solid state is violated by pairs of particles that have 5, respectively 7 next neighbors instead of 6. A preliminary analysis of domain structure was done using the $\Psi_6$ bond order parameter. However, various shortcomings of the $\Psi_6$ bond order parameter have reported recently \cite{short_bond}: The choice of neighborhood definition has an impact on $\Psi_6$ beyond physical interpretation and its inherent discontinuity leads to a lack of robustness.

In this work verify previous results obtained via counting of 5/7-dislocations and the conventional $\Psi_6$ bond order parameter and extend the $\Psi_6$ bond order analysis to measure the defect number fraction. We proceed in the systematic testing of the FDS theory with unprecedented scrutiny: A morphological analysis of lattice structure is performed via Minkowski tensor methods \cite{short_bond,rev_MT_turk,klatt_1,jammed_spheres_mt4,mt_0295-5075-111-2-24002, klatt_2,mt_PhysRevE.96.011301, klatt_3, boebel_dmx}. Minkowski tensors are a tensorial extension of scalar Minkowski functionals. They form a complete family of additive, motion covariant and continuous morphological measures that are sensitive to non-linear properties. They avoid the ambiguity, robustness and discontinuity issues of the bond order parameters and provide highly sensitive morphological measures with a wide range of applications.

As a first step in this work, it is confirmed that the complex plasma phase transitions are in fact not of KTHNY type. This is due to their long-range scaling behavior in the bond order correlation function $g_6(r)$. Then the hypothesis of a fractal relationship between area and boundary length of crystalline domains is tested. Finally, the predicted scale-free relationship of defect fraction and system energy of the FDS theory is verified.

This paper is structured as follows: In Section \ref{sec-theory} the theoretical foundations of the FDS theory are explained. Also theoretical predictions of the KTHNY theory on the bond order correlation function are briefly reviewed. Section \ref{sec-exp-sim} describes the experiments and simulations that were performed and used to test the FDS theory. In Section \ref{sec-mts}, methods are presented: The traditional methods as the bond order metric and the bond order correlation function are described. Then the state of the art morphological analysis methods are introduced: Voronoi tessellations, Minkowski functionals and Minkowski tensors. Based on this introduction an isotropy measure and a symmetry metric is derived. Also the method to cluster particles into homogeneous, ordered domains is explained, as is the method to calculate the particle kinetic energy. Section \ref{sec-results} presents the results obtained by both traditional analysis and Minkowski tensor analysis of the experimental and simulation data. The long range decay scaling of the bond order correlation function, the fractal relationship for energy and defect fraction and for domain area and boundary length are shown.
Finally, in section \ref{sec-conclusion} results are discussed and conclusions are drawn.

\section{Theory \label{sec-theory}}

\subsection{Fractal Domain Structure (FDS) Theory \label{subsec-theory-scale}}

Experimental work, with complex plasmas as model systems \cite{knapek_rec}, provided evidence that fundamental properties of a two-dimensional phase transition are not consistent with the usually assumed KT process. Rather, the findings support the recently developed FDS theory based on the kinetic theory of Frenkel \cite{13o}. The FDS theory was fist introduced in \cite{12o,15o} and is revisited here. The model describes a scale-free phase transition of a two-dimensional N-particle system when the temperature is varied.

At a given energy $E=k_B T$, the $N$-particle system is divided into $ z = N / \left< N_d \right> $ homogeneous domains. Each domain contains $\left< N_d \right>$ particles on average. The domain boundaries are defined by lattice defects (e.g. pairs of pentagons and septagons). The structural order in the individual domains is assumed to be uncorrelated with other domains in the system.

For a mean particle separation $\Delta$ the mean domain radius $\left< r \right>$ is determined by the domain area, consisting of all unit cell areas in the domain, as $\pi \left< r \right>^2 = \pi \left( \Delta/2 \right)^2 \left( N/z \right) $ as 
\begin{equation}
\left< r \right> = 1/2 \left( N/z \right)^{1/2} \Delta.
\end{equation}
Neglecting the interaction between domains, the interface line energy of the boundaries is $ \left< E \right> = 2 \pi \left< r \right> z \sigma$, with the line tension $\sigma$. Substituting $\left< r \right> $ gives 
\begin{equation}
\left< E \right> =  \pi \Delta \left( N z \right)^{1/2} \sigma.
\end{equation}
Due to the arrangement possibilities of the domain structure, the system entropy increases with the number of domains $z$. The number of possible realizations $P$ of the particles ordering characterizes the measure of disorder. It can be calculated by counting the number of possible realizations to distribute $N$ distinguishable particles on $z$ domains, each containing $\left< N_d \right>$ particles.
At first one can choose $\left< N_d \right>$ distinguishable particles from an ensemble of $N$ particles. Then, $ \left< N_d \right>$ particles are chosen from the remaining $N - \left< N_d \right>$ particles with the number of possibilities $p$,
\begin{equation}
p=\binom{N-\left< N_d \right>}{\left< N_d \right>}.
\end{equation}
Repeating this until all domains are completely occupied gives $P$ as the product of all the independent numbers of possibilities: 
\begin{equation}
P = \sum_{i=0}^{z-1} \binom{N- i \left< N_d \right>}{\left< N_d \right>} = N! / \left[ \left( N/z \right)! \right]^z
\end{equation}  
Using Stirling’s formula for sufficiently large $N$ and $N / z$ yields $P \simeq z^N$ . The entropy is $S = \mathrm{ln} \left(P\right)$ and the mean free Helmholtz energy is accordingly
\begin{equation}
\left< F \right> =  \pi \Delta \left( N z \right)^{1/2} \sigma - N T \, \mathrm{ln} \left( z \right) .
\end{equation}  
Assuming thermodynamic equilibrium at all times, it follows from $\partial \left< F \right> / \partial z  = 0$ that 
\begin{equation}
z = \left( 2 T / \pi \Delta \sigma \right)^2 N. \label{eq-z}
\end{equation}
The scaling nature of the domain structure is now introduced as a hypothesis: 
\begin{equation}
\langle N_d \rangle \Delta^2 B = \left[ \Delta \langle N_s \rangle \right]^{1+\alpha}
\label{eq-fractal}
\end{equation}
with $B$ and $\alpha$ constants depending on the shape of the domains. With the above definition $\alpha =1$ if the domain is  circular, for long narrow domains $\alpha \rightarrow 0$, suggesting $0 < \alpha < 1 $ for fractal domains. Substituting this scaling in Eq. \ref{eq-z}, yields the scaling for the total number of particles in all domain boundaries $N_{T} \equiv  z\langle N_s \rangle$:
\begin{equation}
N_T / N \propto T^{2 \alpha/\left( 1+\alpha \right)} \propto E^{2 \alpha/\left( 1+\alpha \right)}. \label{eq-defectenergy}
\end{equation} 

\subsection{Consequences of KTHNY on the Bond Correlation Function $g_6(r)$\label{subsec-theory-kthny}}
A well accepted theory for phase transitions of two-dimensional systems is the KTHNY theory \cite{02o,03o,04o,05o,06o}. Named after Kosterlitz, Thouless, Halperin, Nelson
and Young, it describes the melting of two-dimensional systems with a continuous, second order, defect-mediated phase transition.

The KTHNY theory makes predictions on the long range decay behavior of the bond correlation function for orientational order $g_6 (r)$. It can be defined as
\begin{equation}
g_6(r) \; = \; \sum_{r-\delta r \leq r < r+\delta r} \langle \Psi^{\ast} (\mathbf{r}) \Psi(0) \rangle ,
\end{equation}
with $\Psi (r) = \mathrm{exp} \left( i \theta (\mathbf{r}) \right)$, where $\theta (\mathbf{r})$ denotes the angle between a nearest neighbor bond at position $\mathbf{r}$ and an arbitrary axis. It measures the correlation between the orientation of nearest neighbor bonds separated by the distance $r$.

The KTHNY theory predicts a two-stage melting scenario with an intermediate phase between the solid and liquid state: The hexatic phase. 
In the solid phase $T < T_{c1}$ all dislocations are bound in pairs. Orientational order is preserved in the long range limit: The bond correlation function $g_6(r)$ approaches a finite constant for large distances \cite{04o}.
At $T_{c1}$ the dislocation pairs start to dissociate and for $T > T_{c1}$ the orientational order persists with a slow power-law decay $g_6 (r) \propto r^{−\eta_6(T )} $ \cite{04o}. This transition if well known as the Kosterlitz-Thouless transition. A second transition was discovered by Halperin and Nelson at the temperature $T_{c2} > T_{c1}$: Here the dislocations break up and form free disclinations. The bond order correlation function decays exponentially $g_6(r) \propto  \mathrm{exp}(-r /\xi_6(T))$ \cite{04o, 05o} Table \ref{tab-kthny} summarizes these predictions.

\begin{table}
\caption{\label{tab-kthny} Consequences of the KTHNY theory on the long range scaling behavior of the bond correlation function $g_6(r)$ in different phase regimes. }
\begin{ruledtabular}
\begin{tabular}{cc| cc}
 phase & &$g_6(r)$ scaling  \\
\hline
\rm liquid& ($T>T_{c2}$)& $g_6(r) \propto  \mathrm{exp}(-r /\xi_6(T))$ \\
\rm hexatic& ($T_{c1}<T<T_{c2}$)& $g_6(r) \propto  r^{-\eta_6(T)}$; $\eta_6 < 0.25$ \\
\rm solid& ($T<T_{c1}$)& $g_6(r)=\mathrm{const},\; \mathrm{const} \neq 0$ \\
\end{tabular}
\end{ruledtabular}
\end{table}

\section{Experiments and Simulation \label{sec-exp-sim}}
\begin{figure}[!tbp]
	\captionsetup[subfigure]{position=top,singlelinecheck=off,labelfont=bf,textfont=normalfont, justification=raggedright, margin=10pt,captionskip=-1pt,labelformat=empty }
	\hspace*{-0.2cm} 
	\subfloat[\label{sub_exp1}]{%
	\includegraphics[width=0.48\textwidth]{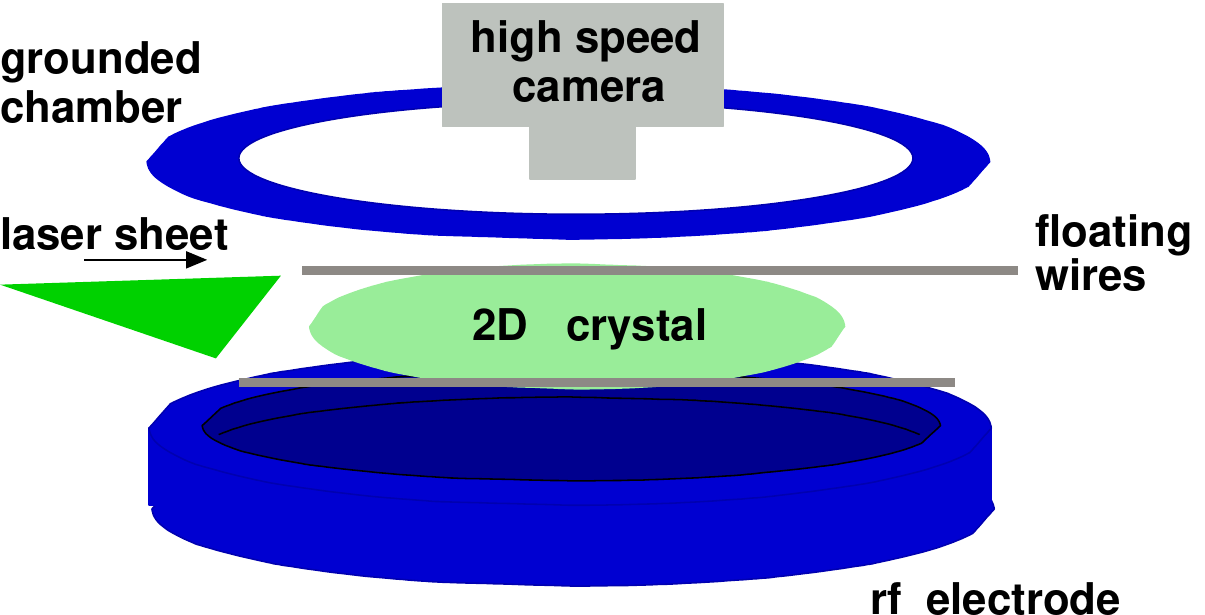}}
\caption{\label{fig-exp1} \small Sketch of the experimental setup used for the presented crystallization experiments \cite{15o}. A two-dimensional crystal is levitated in the plasma sheath region above the lower rf electrode. A glass window in the upper chamber flange provides optical access for a high speed camera from the top viewpoint. Particles are illuminated by a vertically thin, horizontally spread laser sheet. Two wires are mounted inside the chamber for electric particle manipulation. These are normally floating, but can be fed with a short electric pulse to melt the particle system.}
\end{figure}

To study the phase transition in a genuine two-dimensional system, experiments \cite{knapek_rec} were performed with two-dimensional complex plasmas: many-particle systems consisting of electrons, ions, neutral gas atoms, and charged micrometer sized particles. A sketch of the experimental setup is provided in Fig. \ref{fig-exp1}. An example image of an experimental data set is shown in Fig. \ref{fig-exp2}. Movies for all data sets are provided in the supplemental material \cite{supp_rec}.

Melamine-formaldehyde particles with a diameter of 9.19 $\mu$m and a mass of $6.14\times10^{-13}$kg were injected into an argon radio-frequency (rf) discharge ignited between a horizontal, capacitively coupled electrode mounted on the bottom of a vacuum chamber, and the grounded chamber walls. Due to the balance of electron- and ion-streams onto their surface, the particles acquired a negative charge. The electric fields in the plasma sheath region above the electrode then levitated particles against gravity (usually several mm above the electrode surface). Additionally, an elevated rim on the electrode provided a radial confinement by shaping the electric potential inside the chamber. The injected particles then formed a crystalline single layer with a hexagonal crystal structure, which could temporarily be destroyed by applying a negative electric pulse (duration: $0.2$~s, amplitude: $-250$~V) to two parallel wires ($58.7$~mm apart from each other) mounted at approximately the levitation height of the mono-layer. The particle system then was left to recrystallize under constant "environmental" conditions, i.e. pressure and rf-power.

To obtain particle trajectories, the particle layer was illuminated by a $532$~nm Nd:YAG laser, adjusted to provide a vertically thin, horizontally spread sheet of light. The light reflected  by the particles was then observed by a high-speed camera with a frame rate of $250$ frames per second (fps) and a spatial resolution of $0.03$~mm/px from the top viewpoint through a glass window. To reduce the effect of pixel noise during the image analysis, each two consecutive images were later averaged, yielding an effective frame rate of $125$~fps \cite{15o}. The number of particles in the field of view of the camera was approximately $2000$, which amounts to approximately $10-15$~\% of the total number of particles in the mono-layer.

Experiments were performed at 11 different plasma conditions: the neutral gas pressure was varied between $1.15-2.3$~Pa and the peak-to-peak rf voltages $U_{PP}$ at the electrode were chosen in the range $[-134,-214]$~V.

Additionally, another data set from \cite{14o} was included in the analysis. Here, the gas pressure was $1.94$~Pa, $U_{PP}$ was $-172$~V, the recording frame rate was $500$~fps (effective frame rate after averaging each 3 consecutive images: $166.667$~fps), and the spatial resolution was $0.034$~mm/px.

Details of the experimental setup are given in \cite{14o,15o}.

To complement the experimental results, the outcome of a molecular dynamics simulation of the crystallization of a mono-layer of 3000 particles in a parabolic confinement is presented in addition. The simulation parameters were chosen to meet the experimental conditions: the damping rate was $2$~Hz, the time step $0.01$~s, particle mass and charge were $6.1\times10^{-13}$~kg and $-12000$~e, respectively. The particles were initially heated to $230$~eV, and then allowed to cool until they reached a crystalline state. The parabolic potential used in the simulation gives rise to deviations from the experiments. The confinement in the experiments is non-parabolic due to the presence of the electrodes used to induce the electric shock causing the melting. The non-parabolic confinement in the experiments leads to a constant particle density whereas the parabolic confinement in the simulation gives rise to a radially decreasing particle density. Also the expansion before melting and relaxation during crystallization of the system is affected by the difference in the confinement potential. Details of the simulation procedure are given in \cite{16o}.

Here, the results of these earlier experiments and simulation are analyzed employing Minkowski tensor methods and compared with previous results. The particular parameters of each experiment are given in Table \ref{tab-exp}.

\begin{table}
\caption{\label{tab-exp}Parameters of the experiments and the simulation. Neutral gas pressure $p$, Epstein damping coefficient $\nu$, peak-to-peak rf voltage $U_{PP}$ at the driven electrode and the mean particle separation $\Delta$ obtained from the pair correlation functions. The Epstein damping coefficient $\nu$, a measure for the damping rate of the particle motion due to scattering on neutral gas atoms, was calculated from the discharge parameters given in Ref. \citep{17o}, using the reflection index $\delta = 1.26$ as measured in
 Ref. \citep{18o}.}
\begin{ruledtabular}
\begin{tabular}{ccccccc}
 &$p (\mathrm{Pa})$ &$\nu (\mathrm{Hz}) $ &$U_{PP} (\mathrm{V}) $
 &$\Delta (\mathrm{mm}) $  \\
\hline
\rm I& 1.93 & 2.27 & -138 &0.60\\
\rm II& 1.36 & 1.60 & -144 &0.61 \\
\rm III & 2.29 & 2.69 & -134 &0.61 \\
\rm IV & 1.15 & 1.35 & -184 &0.60 \\
\rm V & 1.36 & 1.60 & -180 &0.60 \\
\rm VI & 1.68 & 1.97 & -176 &0.60 \\
\rm VII & 2.12 & 2.49 & -172 &0.60 \\
\rm VIII & 2.30 & 2.70 & -172 &0.60 \\
\rm IX & 1.36 & 1.60 & -214 &0.57 \\
\rm X & 1.93 & 2.27 & -206 &0.51 \\
\rm XI & 2.30 & 2.70 & -200 &0.53 \\
\rm XII & 1.94 & 2.28 & -172 &0.59 \\
\rm Simulation (S) & ... & 2 & ... &0.8 \\
\end{tabular}
\end{ruledtabular}
\end{table}

\begin{figure}[!tbp]
	\captionsetup[subfigure]{position=top,singlelinecheck=off,labelfont=bf,textfont=normalfont, justification=raggedright, margin=10pt,captionskip=-1pt,labelformat=empty }
	\hspace*{-0.2cm} 
	\subfloat[\label{sub_exp2}]{%
	\includegraphics[width=0.48\textwidth]{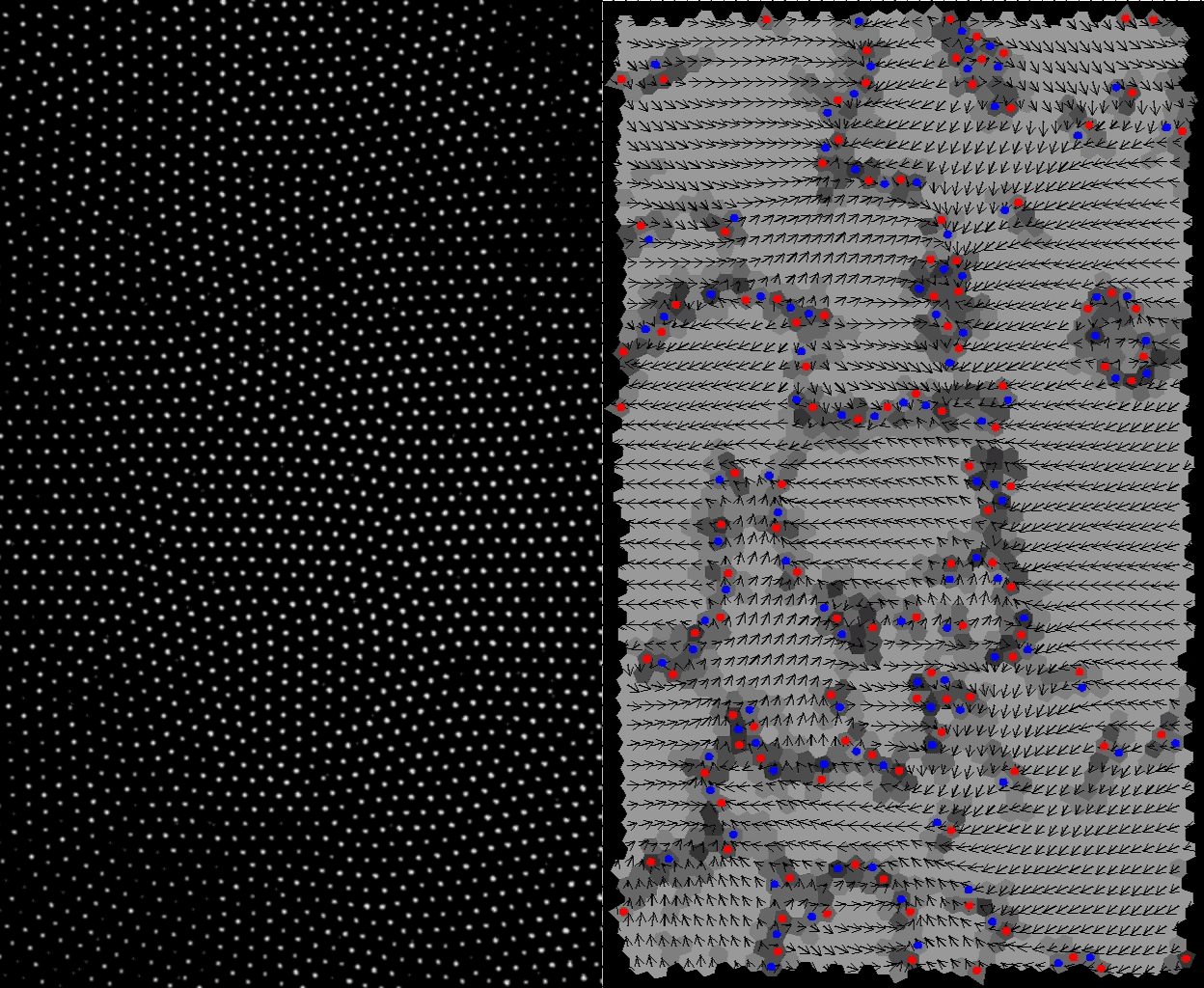}}
\caption{\label{fig-exp2} \small Left: Image of a two-dimensional plasma crystal of the experimental data set XII for time $t=6.0$ s. The field of view is $18$ mm x $25$ mm.  Right: Greyscale plot of the $\Psi_6$ bond order parameter. Dark Voronoi cells have low $\Psi_6$ values. The direction of the argument of $\Psi_6$ is indicated with arrows. Dislocations with 5, respectively 7 neighbours are marked with red, respectivly blue dots.}
\end{figure}	
\section{Methods \label{sec-mts}}
The bond order parameters $\Psi_6$ were introduced in 1983 \cite{psi_stein} and quickly became a standard tool to quantify crystalline structures. However, recent work \cite{short_bond} has shown that the calculation of $\Psi_6$ has some conceptual drawbacks. The choice of neighborhood definition causes an ambiguity of $\Psi_6$ beyond physical interpretation and its inherent discontinuity leads to a lack of robustness.

On the other hand the Minkowski functionals are a continuous and robust tool for morphological data analysis, known since the early 20th century \cite{Minkowski1903}. Only recently the hierarchy of Minkowski valuation was extended to tensor valued quantities called Minkowski tensors \cite{mink_alesker}. Minkowski functionals and tensors are sensitive to any $n$-point correlation function and thus can give new insights to processes beyond the capability of conventional (linear) methods, e.g. $\Psi_6$, $g(r)$ or $g_6(r)$. 
A commonly used method for quantifying the local structure of points (or discs) is by construction of a nearest neighbor network on which quantitative structure metrics are computed (e.g. $\Psi_6$). The ambiguity of the neighborhood selection can be circumvented by using the method of the bijective Voronoi tessellation, based on the idea of a Wigner-Seitz cell for each particle.

In the following, the $\Psi_6$ bond order metric, the bond orientational correlation function $g_6(r)$, Voronoi tessellation, Minkowski functionals and tensors, as well as a clustering algorithm via DBSCAN and the energy calculation via velocity distribution fits, are introduced as methods used throughout this paper.

\subsection{Bond Order Parameter $\Psi_6$ \label{subsec-psi6}}
\begin{figure}[!tbp]
	\captionsetup[subfigure]{position=top,singlelinecheck=off,labelfont=bf,textfont=normalfont, justification=raggedright, margin=10pt,captionskip=-1pt }
	\hspace*{-0.2cm} 
	\subfloat[]{%
	\includegraphics[width=0.24\textwidth]{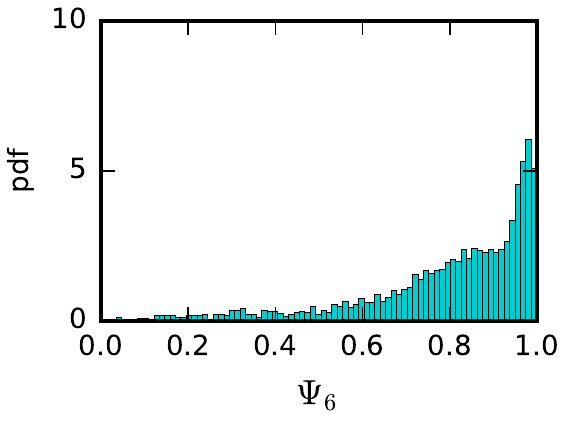}}
	\hspace*{-0.1cm} 
	\subfloat[]{%
	\includegraphics[width=0.24\textwidth]{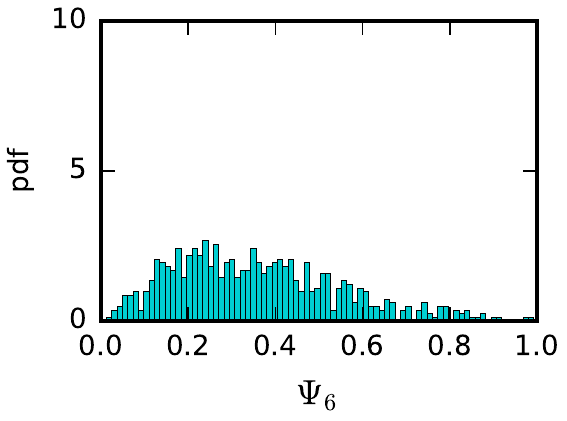}}
	\vspace{-0.5cm} 
	\hspace*{-0.2cm}
	\subfloat[]{%
	\includegraphics[width=0.24\textwidth]{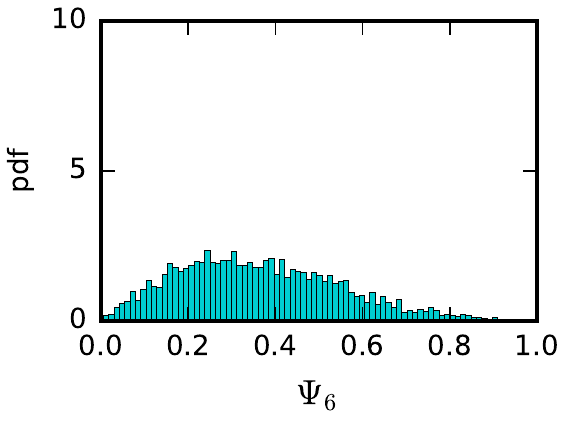}}
	\hspace*{-0.1cm} 
	\subfloat[]{%
	\includegraphics[width=0.24\textwidth]{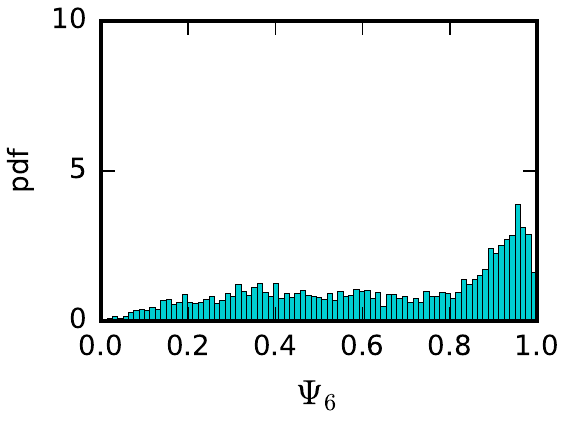}}
	\vspace{-0.5cm} 
	\hspace*{-0.2cm}
	\subfloat[]{%
	\includegraphics[width=0.24\textwidth]{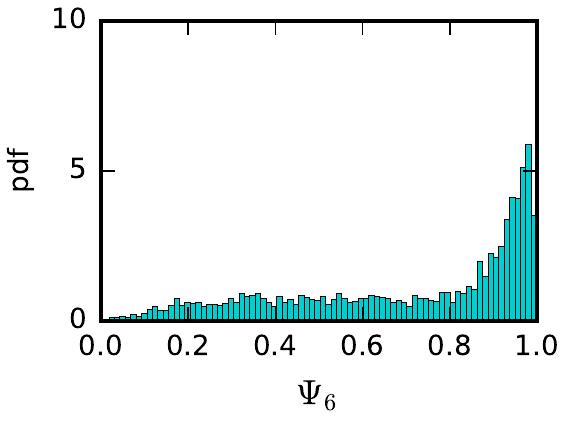}}
	\hspace*{-0.1cm} 
	\subfloat[]{%
	\includegraphics[width=0.24\textwidth]{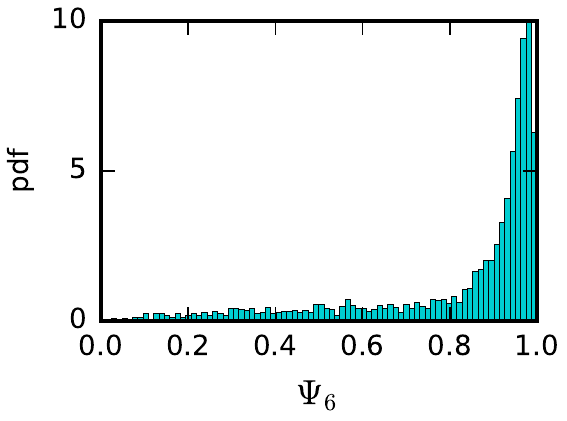}}
\caption{\label{fig-histp6} \small Typical histograms for the $\Psi_\mathrm{6}$ measure as time $t$ evolves representatively shown for experiment X (see Table \ref{tab-exp}). (a) $t=3.00 $ s: Before melting a large peak for $\Psi_\mathrm{6}$ values close to $\Psi_\mathrm{6}=1$ is a signature of the crystalline state. (b)-(c) $t=3.30$ s, $t=4.50$ s: The distribution broadens after melting. (d)-(e)  $t=6.00$ s, $t=7.20$ s: During recrystallization the distribution shifts to larger $\Psi_\mathrm{6}$ values. (f) $t=12.00$ s: For late times the large peak for values close to $\Psi_\mathrm{6}=1$ is recovered in the recrystallized state.}
\end{figure}

In a fist step of a thorough investigation of the domain structure, irregular lattice sites are identified as defects via the $\Psi_6$ bond order parameter \cite{psi_stein}. It is defined as
\begin{equation}
\Psi_{6} = 1/n_k \times \sum_{m=1}^{n_k} \mathrm{exp} \left( 6 i \Theta_{km} \right)
\end{equation}

for each lattice site $k$. Here, $n_k$ is the number of nearest neighbors of particle $k$, $\Theta_{km}$ is the angle of the bond between particles $k$ and $m$ to an arbitrary chosen axis (we chose the x-axis), and $i$ is the imaginary unit. For hexagonal ordered lattice sites, the modulus $| \Psi_{6} |$ is close to $1$, whereas it tends to zero for distorted ones and therefore also for defects. In order to distinguish ordered from disordered sections, a cut off value of $\Psi_{\mathrm{6},\mathrm{thresh}} > \Psi_{\mathrm{6},\mathrm{defects}}$ (corresponding to Voronoi cells close to $| \Psi_\mathrm{6} |=1$ interpreted as in the crystalline state) is chosen. The specific value that is chosen for $\Psi_{\mathrm{6},\mathrm{thresh}}$ is indicated in each case in the result section \ref{sec-results}. The number fraction of these particles identified as in the crystalline state will be referred to as $\Psi_\mathrm{6}$ measure in the following. Typical histograms for $\Psi_6$ for the analyzed recrystallization processes are shown in Fig. \ref{fig-histp6} for increasing time steps.
\subsection{Bond Correlation Function $g_6(r)$ \label{subsec-g6}}
The bond correlation function $g_6 (r)$ for the orientational order \ref{subsec-theory-kthny} is calculated via:
\begin{equation}
g_6(r)  =  \left| \dfrac{1}{N_B} \sum_{l=1}^{N_B} \dfrac{1}{n(l)} \sum_{k=1}^{n(l)} \mathrm{exp} \left\lbrace 6i \left( \theta (r_k) - \theta(r_l) \right) \right\rbrace  \right|
\end{equation}
Here, $N_B$ is the total number of bonds in the crystal, $n(l)$ is the number of bonds at distance $r$ from bond $l$, $\theta_i$ the angle of bond $i$ at $r_i$ to an arbitrary
axis. For a perfect hexagon $g_6 (r) \equiv 1$. Since we are only interested in the long range decay, and not the exact shape of $g_6$ with its peaks in the close range regime, we choose large bins, i.e. large values of $n(l)$.

For a solid crystalline state $g_6(r)$ should be constant and close to $1$ \cite{04o}. However, for the plasma crystal data sets analyzed here, we find $g_6(r)$ to be a linearly decaying function. This is because the crystal is made up of homogeneous domains, separated by defect lines, whose structural order is uncorrelated with neighboring domains.
Since power-law, respectively exponential decay is predicted in hexatic, respectively liquid states \cite{04o, 05o} for large $r$, following models are fitted to the experimental and simulation data sets: (a) linear decay $g_6(r)=A_1 + c_6 \cdot r$, (b) exponential decay $g_6(r)=A_2 \cdot \mathrm{exp}(-r/\xi_6)$ and (c) power-law decay $g_6(r)=A_3 \cdot r^{-\eta_6}$. To determine the best model the goodness of best fits is compared using the chi-squared $\chi^2$ statistic. Lower values indicate a higher goodness of fit. This method was already applied in order to test for a hexatic phase \cite{15o,chi2_PhysRevE.75.031402}.
\subsection{Voronoi Tessellation \label{subsec-voronoi}}

An approach for quantifying local structure is provided by the analysis of the Voronoi diagram. The Voronoi diagram is the partition of space into the same number of convex cells as there are discs in the packing. The Voronoi cell of each disc is the region of space closer to that given disc than to any other disc. For the special case of three- or two-dimensional crystal lattices the Voronoi cell is called Wigner-Seitz cell. In the field of granular matter, Voronoi diagrams have been used to determine distributions of local packing fractions \cite{voronoi6_PhysRevLett.96.018002, voronoi7_0295-5075-79-2-24003, voronoi8_PhysRevE.77.021309}, spatial correlations \cite{voronoi9_0295-5075-97-3-34004} and correlations with particle motion \cite{voronoi10_PhysRevLett.101.258001}.

Recently, studies provided insight into the local structure of sphere packings and sphere ensembles by analyzing the shape of Voronoi cells, in particular their degree of anisotropy or elongation \cite{voronoi11_0295-5075-90-3-34001, voronoi12_1742-5468-2010-11-P11010, voronoi13_kapfer2012a, voronoi}.

Here, the structure of the Voronoi tessellation, obtained from particle positions, is analyzed using Minkowski functional and tensor methods. The boundary particles were discarded from the analysis since they have no neighboring particles needed to define their Voronoi cells.
\subsection{Minkowski Functionals \label{subsec-mf}}
For a body $K$ with a smooth boundary contour $\partial K$ embedded in $D$-dimensional euclidean space the $D+1$ Minkowski functionals are, up to constant factors, defined as:
\begin{equation}
\begin{aligned}
W_0(K) \; &= \; \int_K \, \mathrm{d}^D r
\\
W_{\nu} (K) \; &= \; \int_{\partial K} \, G_{\nu}(r) \, \mathrm{d}^{D-1} r \quad , \quad 1 \leq \nu \leq D
\end{aligned}
\label{eq-mf}
\end{equation}
$G_{\nu}(r)$ are the elementary symmetric polynomials of the local principal curvatures as defined in differential geometry.

In two-dimensional euclidean space the Minkowski functionals, up to constant factors, are $W_0(K)$ (area), $W_1(K)$ (circumference) and $W_2(K)$ (euler characteristic):\begin{equation}
\begin{aligned}
W_0(K) &= \int_K \mathrm{d}^2 r 
\\
W_1(K) &= \int_{\partial K} \mathrm{d} r 
\\
W_2(K) &=\int_{\partial K}  \kappa (r) \: \mathrm{d} r 
\end{aligned}
\label{eq-mf3d}
\end{equation}
Here, $\kappa (r)$ is the local curvature.

Minkowski functionals are motion invariant, additive and conditionally continuous. They form a complete family of morphological measures. Or vice versa: Any motion invariant, (conditionally) continuous and additive functional is a superposition of the (countably many) Minkowski functionals. They are nonlinear measures sensitive to higher order correlations. Applications are e.g. curvature energy of membranes \cite{mink_app1}, order parameter in Turing patterns \cite{mink_app2_Mecke_turing}, density functional theory for fluids (as hard balls or ellipsoids) \cite{mink_app3_density_func, mink_app_mecke}, testing point distributions (find clusters, filaments, underlying point-process) or searching for non-Gaussian signatures in the CMB \cite{schmalz,WINITZKI199875, modest_1, modest_2,modest_3}.

\subsection{Minkowski Tensors \label{subsec-mt}}

In order to also account for directional properties it is natural to extend the scalar valued Minkowski functionals to tensor valued quantities called Minkowski tensors. Applications of Minkowski tensors range from the analysis of cellular, granular and porous structures to the classification of crystal types \cite{klatt_1,jammed_spheres_mt4,mt_0295-5075-111-2-24002, klatt_2,mt_PhysRevE.96.011301, klatt_3, boebel_dmx}. They are defined as \citep{rev_MT_turk}:
\begin{equation}
\begin{aligned}
W_0^{a,0}(K) \; &:= \; \int_K \, \mathrm{d}^D r \; \; \mathbf{r}^{\odot a}
\\
W_{\nu}^{a,b} (K) \; &:= \; 1/D  \int_{\partial K}  \mathrm{d}^{D-1} r  \; \; G_{\nu}(r) \; \; \mathbf{r}^{\odot a} \odot \mathbf{n}^{\odot b}
\end{aligned} 
\label{eq-mt}
\end{equation}
Here, $\odot$ denotes the symmetric tensor product $x \odot y = 1/2 \, (x \otimes y + y \otimes x)$. Again $G_{\nu}(r)$ are the elementary symmetric polynomials of the local principal curvatures as defined in differential geometry. $a$ counts the number of position vectors $\mathbf{r}$, $b$ counts the number of normal vectors $\mathbf{n}$ in the tensor product. Thus the rank of each tensor is the tuple $(a,b)$.

Similar to Minkowski functionals the attractiveness of Minkowski tensors is due to their manifold applications. Further, they are founded on a solid mathematical framework: A strong completeness theorem by Alesker \cite{mink_alesker} states that all morphological information that is relevant for additive properties of a body $K$ is represented by the Minkowski tensors. Any motion covariant, conditionally continuous and additive tensor valued functional is a superposition of the (countably many) Minkowski tensors.
\begin{figure}[tbp]
	\captionsetup[subfigure]{position=top,singlelinecheck=off,labelfont=bf,textfont=normalfont, justification=raggedright, margin=10pt,captionskip=-1pt,labelformat=empty }
	\hspace*{-0.2cm} 
	\subfloat[]{%
	\includegraphics[width=0.45\textwidth]{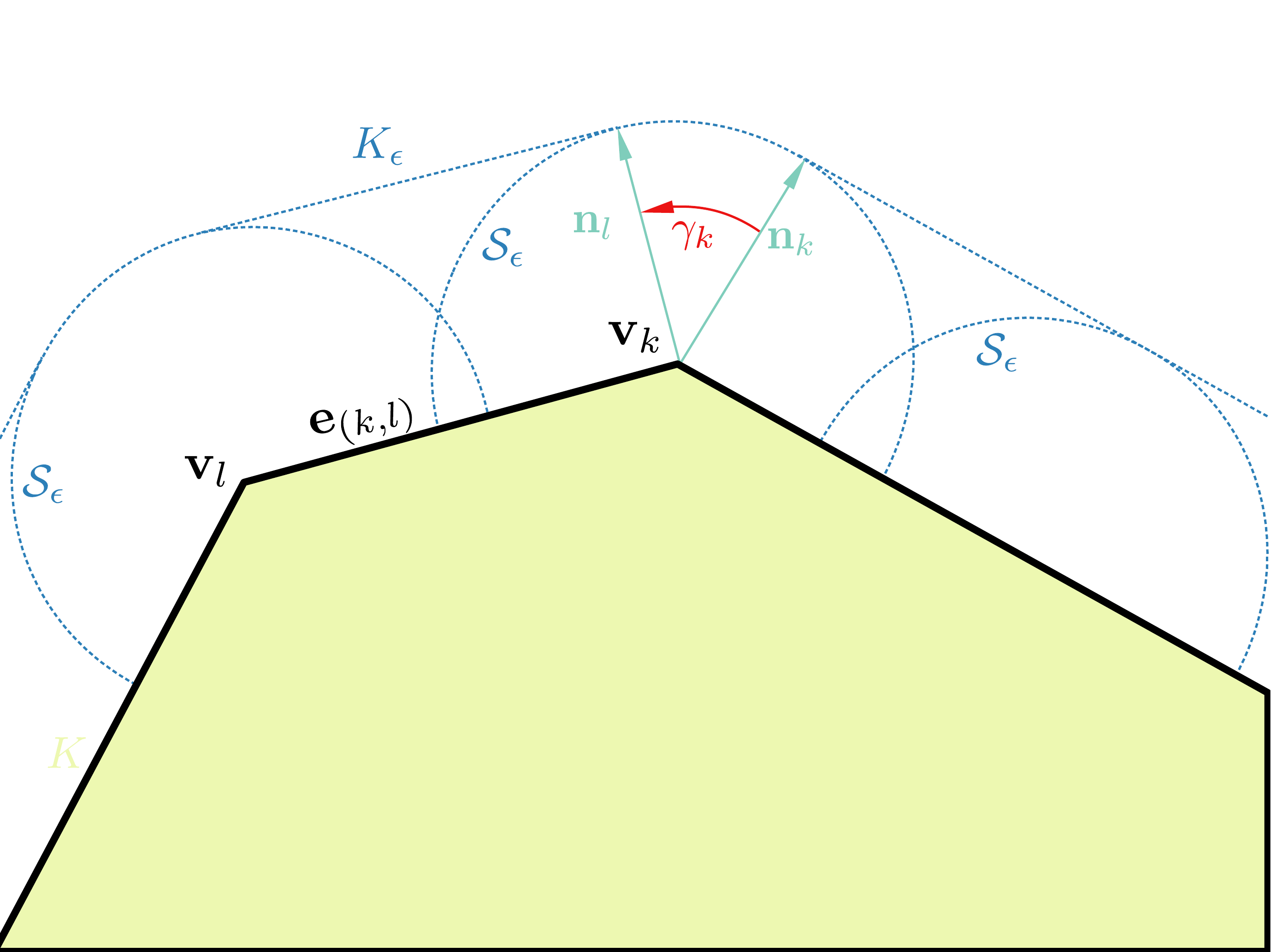}}
\caption{\label{fig-mtcalc} \small Illustration for the explicit calculation of Minkowski tensors of a body $K$ via $K_{\epsilon}$.}
\end{figure}	

The Minkowski tensors are defined as curvature integrals over smooth boundary surfaces. In order to calculate them for polygonal bodies $P$ we consider the parallel body construction $P_{\epsilon}= P \uplus S_{\epsilon}$ \cite{mt2d}. $S_{\epsilon}$ is a disk of radius $\epsilon > 0$ and $\uplus$ is the Minkowski sum (defined as: $K_1 \uplus K_2 = \left\lbrace p_1 + p_2 \mid p_1 \in K_1, p_2 \in K_2 \right\rbrace$). Thus, $P_{\epsilon}$ is the union of all disks $S_{\epsilon}$ with origins at all points in $P$, illustrated in Fig. \ref{fig-mtcalc}. Performing the limit $\epsilon \to 0 $ then yields the tensor $W_{\nu}^{a,b} (P) =  \lim_{\epsilon \to 0} W_{\nu}^{a,b} (P_{\epsilon}) $. Consider the polygonal representation of $P$ by its vertices $\mathbf{v}_k$. Then the edges between vertices $\mathbf{v}_k$ and $\mathbf{v}_l$ are $\mathbf{e}_{(k,l)}=\mathbf{v}_l-\mathbf{v}_k$ with normal vectors $\mathbf{n}_{(k,l)}= R \, \mathbf{e}_{(k,l)} / \left| \mathbf{e}_{(k,l)} \right|$. $R = \bigl( \begin{smallmatrix}0 & -1\\ -1 & 0\end{smallmatrix}\bigr)$ is the $\pi / 2$ rotation matrix. $\gamma_k$ is the angle between $\mathbf{n}_{(k-1,k)}$ and $\mathbf{n}_{(k,k+1)}$. Using these definitions we can obtain the explicit formula. Here we present formulae for the second rank in position vectors circumference (Eq. \ref{eq_circtens}) and the second rank in normal vectors euler tensors (Eq. \ref{eq_eulertens}) as examples. $E$ is the unit matrix.

\begin{equation}
\begin{aligned}
&W_{1}^{2,0}(P) \, = \, \lim_{\epsilon \to 0} \dfrac{1}{2}  \int_{\partial P_{\epsilon}}  \mathrm{d} r  \; \; \mathbf{r} \odot \mathbf{r} \\
&= \, \dfrac{1}{6} \sum_{(k,l)}  \left| \mathbf{e}_{(k,l)} \right| \cdot \\
&\left( \begin{matrix} v_{kx}^2 + v_{kx} v_{lx} + v_{lx}^2  & v_{kx}v_{ky} + v_{kx} v_{ly} + v_{lx}v_{ly}\\ v_{ky}v_{kx} + v_{ky} v_{lx} + v_{ly}v_{lx} & v_{ky}^2 + v_{ky} v_{ly} + v_{ly}^2 \end{matrix} \right) \\
\end{aligned}
\label{eq_circtens}
\end{equation}

\begin{equation}
W_{2}^{0,2}(P)\, = \, \lim_{\epsilon \to 0} \dfrac{1}{2}  \int_{\partial P_{\epsilon}}  \mathrm{d} r  \; \; \kappa(r) \;\mathbf{n} \odot \mathbf{n} \, = \, 4 \; W_2 \; E.
\label{eq_eulertens}
\end{equation}

A list of expressions for two-dimensional tensors up to rank two is available in \cite{mt2d}.

\subsection{MT2 Isotropy Index \label{subsec-beta}}
For a body $K$ and each second rank Minkowski tensor $W_{\nu}^{a,b}(K)$ an isotropy index $\beta$ can be defined as the ratio between the smallest and largest eigenvalue $\lambda_{min}$ and $\lambda_{max}$ of the $D \times D$-matrix representing each Minkowski tensor: \cite{rev_MT_turk}
\begin{equation}
\beta_{\nu}^{a,b}(K) : = \dfrac{\lambda_{min} \left( W_{\nu}^{a,b}(K) \right) }{\lambda_{max} \left( W_{\nu}^{a,b}(K) \right)}
\label{eq-beta}
\end{equation}
The dimensionless isotropy index is a pure shape measure. It is invariant under isotropic scaling of $K$. For example in two dimensions the isotropy index $ \beta =1 $ is obtained for a circle or a square. For a rectangle one obtains $ \beta = \mathrm{shorter}/\mathrm{longer} \, \mathrm{edge}$. Thus this isotropy index is an isotropy measure only in the sense of elongation. $\beta$ provides equivalent information as the improved, area weighted bond order metric proposed in \cite{short_bond}.

\begin{figure}[!tbp]
	\captionsetup[subfigure]{position=top,singlelinecheck=off,labelfont=bf,textfont=normalfont, justification=raggedright, margin=10pt,captionskip=-1pt }
	\hspace*{-0.2cm} 
	\subfloat[]{%
	\includegraphics[width=0.24\textwidth]{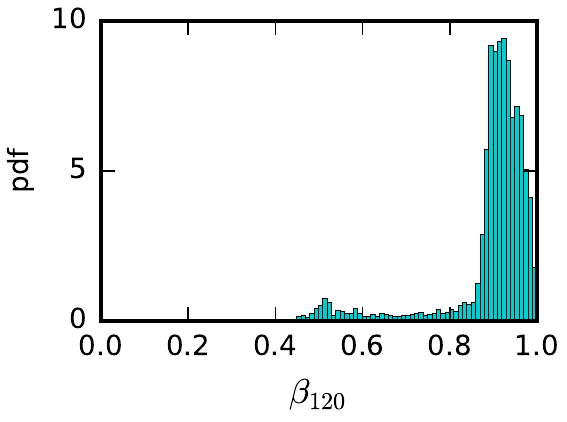}}
	\hspace*{-0.1cm} 
	\subfloat[]{%
	\includegraphics[width=0.24\textwidth]{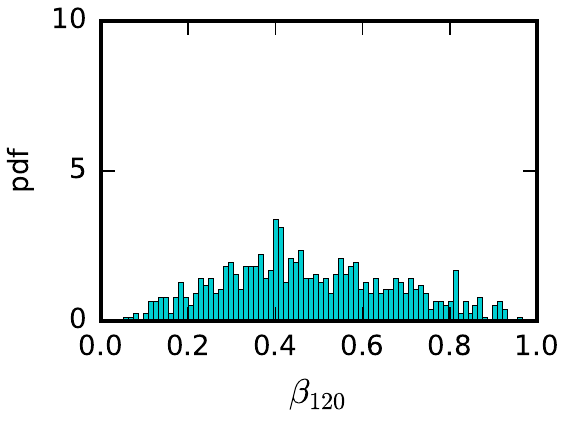}}
	\vspace{-0.5cm} 
	\hspace*{-0.2cm}
	\subfloat[]{%
	\includegraphics[width=0.24\textwidth]{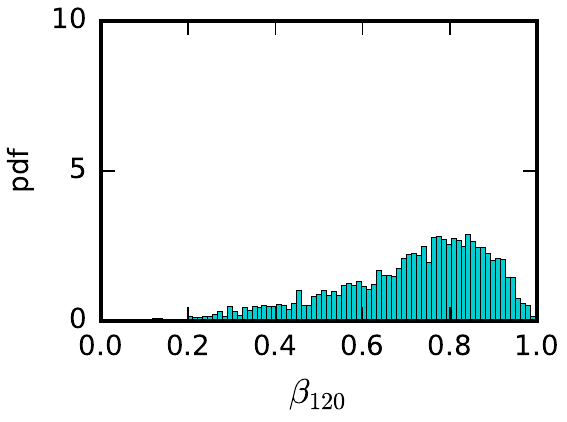}}
	\hspace*{-0.1cm} 
	\subfloat[]{%
	\includegraphics[width=0.24\textwidth]{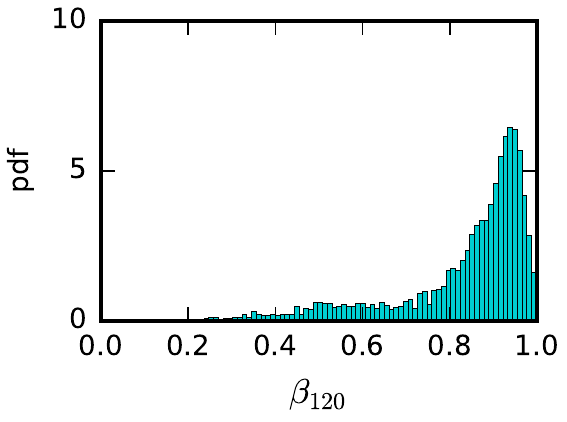}}
	\vspace{-0.5cm}
	\hspace*{-0.2cm}
	\subfloat[]{%
	\includegraphics[width=0.24\textwidth]{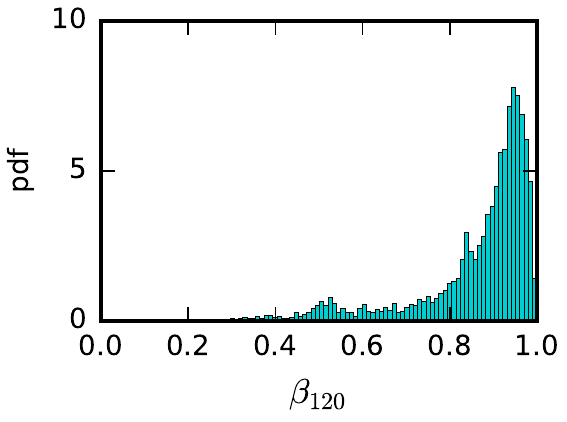}}
	\hspace*{-0.1cm} 
	\subfloat[]{%
	\includegraphics[width=0.24\textwidth]{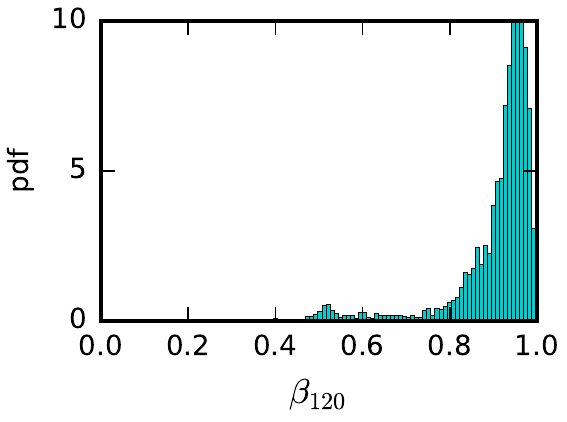}}
\caption{\label{fig-betahist} \small Typical histograms for the $\mathrm{MT2}$ measure as time $t$ evolves representatively shown for experiment X (see Table \ref{tab-exp}). (a)$t=3.00$ s: Before melting a large peak for $\beta$ values close to $\beta=1$ is a signature of the crystalline state. (b) $t=3.30$ s The distribution broadens after melting. (c)-(e) $t=4.50$ s, $t=6.00$ s, $t=7.20$ s: During recrystallization the distribution shifts to larger $\beta$ values. (f) $t=12.00$ s: For late times the large peak for $\beta$ values close to $\beta=1$ is recovered in the recrystallized state.}
\end{figure}

The rank two Minkowski tensor analysis carried out in this study is done by calculating the isotropy index $\beta$ Eq. (\ref{eq-beta}) locally for every Voronoi cell and for every time step in the experimental and simulation data using the circumference Minkowski tensor $W_{1}^{2,0}$, since it is most sensitive to changes of elongation of Voronoi cells. In order to distinguish ordered from disordered Voronoi cells a cut-off value of $\beta_{\mathrm{thresh}} > \beta_{\mathrm{defects}}$ (corresponding to isotropic Voronoi cells interpreted as in the crystalline state) is chosen. The specific value that is chosen for $\beta_{\mathrm{thresh}}$ is indicated in each case in the result section \ref{sec-results}. The number fraction of these particles identified as in the crystalline state will hereinafter be referred to as $\mathrm{MT2}$ measure. Typical histograms for $\beta$ for the analyzed recrystallization processes are shown in Fig. \ref{fig-betahist} as time evolves.
\subsection{MT4 Symmetry Metric \label{subsec-delta}}

In order to distinguish between structures of high symmetry, i.e. differentiate between crystalline structures (hcp, fcc, etc.), higher ranked tensors have to be applied. For rank four and higher, isotropic symmetry is distinct from cubic symmetry. (This is evidenced by the appearance of a second independent shear modulus when transitioning from isotropic to cubic symmetry in the theory of linear elasticity, which is formulated using a rank-four tensor \cite{walpole_elastic_theory}.) This method has been used in hard sphere systems to characterize random close packings \cite{jammed_spheres_mt4}.

For brevity, only the simplest rank four Minkowski tensor is considered: 
\begin{equation}
W_1^{0,4} (K) = \dfrac{1}{2} \int_{\partial K} \mathrm{d} r \; \mathbf{n}(\mathbf{r}) \otimes \mathbf{n}(\mathbf{r}) \otimes \mathbf{n}(\mathbf{r}) \otimes \mathbf{n}(\mathbf{r}).
\end{equation}
In the polygonal representation its components, labeled $\mu, \nu, \tau, \sigma \in (x,y)$ are:
\begin{equation}
\left[ W_1^{0,4} (P) \right]^{\mu \nu \tau \sigma} = \dfrac{1}{2}  \sum_{(k,l)} \left| \mathbf{e}_{(k,l)} \right| \cdot n_{(k,l)}^{\mu}n_{(k,l)}^{\nu}n_{(k,l)}^{\tau}n_{(k,l)}^{\sigma}.
\end{equation}
Since it is translation invariant and symmetric (i.e. it holds for the components $\left[ W_1^{0,4} \right]^{\mu \nu \tau \sigma} = \left[ W_1^{0,4} \right]^{(\mu \nu \tau \sigma)}$ ) it has, in two dimensions, only $5$ independent elements instead of $16$. The round brackets denote cyclic permutation.

A morphological metric suitable for characterizing systems of spherical particles should be rotationally invariant since the physics does not a priori designate a preferred direction. Thus, the tensor $W_1^{04}$ should not be directly used. Instead, rotational invariants are constructed \cite{Kapfer2012}. This is done by borrowing ideas from the theory of the elastic stiffness tensor.

The tensor $W_1^{0,4} (K)$ is rewritten in the Mehrabadi supermatrix notation \cite{MEHRABADI} as a $3\times3$ matrix:
\begin{equation}
M=\begin{bmatrix} 
S_{xxxx} & S_{xxyy} & S_{xxzz} \\
S_{yyxx} & S_{yyyy} & \sqrt{2} \, S_{yyxy} \\
\sqrt{2} \, S_{xyxx} & \sqrt{2} \, S_{xyyy} & 2 S_{xyxy} \\ 
\end{bmatrix}
\end{equation}
where $S=W_1^{0,4} (K) /W_1 (K)$.

Then, the three-tuple formed by the eigenvalues $\zeta_i$ of $M$ (in descending order) may be considered a symmetry fingerprint of the polyhedron $K$. It is invariant under rotation, scaling and translation of the polyhedron $K$.
Using the signature eigenvalue tuple $\zeta_i$  of $M$ it is possible to define a distance measure on the space of bodies induced by the Euclidean distance:
\begin{equation}
\Delta (K_1, K_2): =  \left( \sum_{i=1}^6 \left( \zeta_i(K_1) - \zeta_i(K_2)  \right)^{2}  \right)^{1/2}.
\label{eq-delta}
\end{equation}
$\Delta (K_1, K_2)$ is a pseudometric. It is positive definite, symmetric, the triangle inequality holds, however, the coincidence axiom $ \Delta(K_1,K_2) = 0 \Leftarrow K_1=K_2$ is only an implication and not an equivalence. For example $ \Delta (\mathrm{sphere}, \mathrm{dodecahedron}) = 0$. To distinguish dodecahedra from spheres one needs to employ even higher rank tensors. 

\begin{figure}[!tbp]
	\captionsetup[subfigure]{position=top,singlelinecheck=off,labelfont=bf,textfont=normalfont, justification=raggedright, margin=10pt,captionskip=-1pt }
	\hspace*{-0.2cm} 
	\subfloat[]{%
	\includegraphics[width=0.24\textwidth]{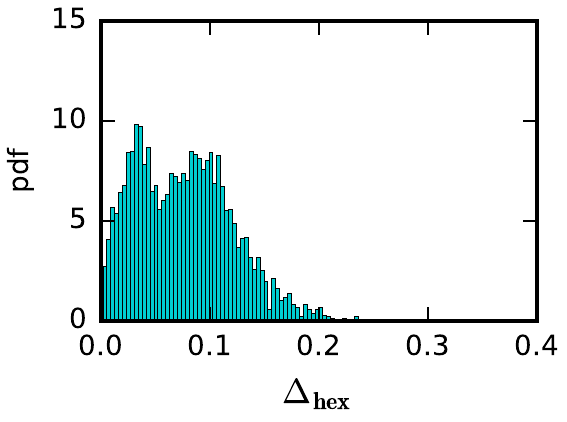}}
	\hspace*{-0.1cm} 
	\subfloat[]{%
	\includegraphics[width=0.24\textwidth]{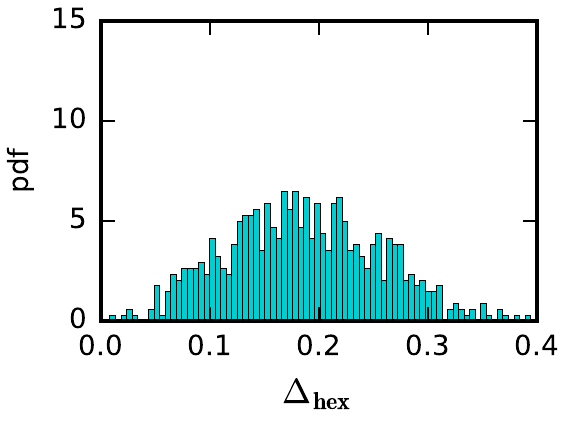}}
	\vspace{-0.5cm} 
	\hspace*{-0.2cm}
	\subfloat[]{%
	\includegraphics[width=0.24\textwidth]{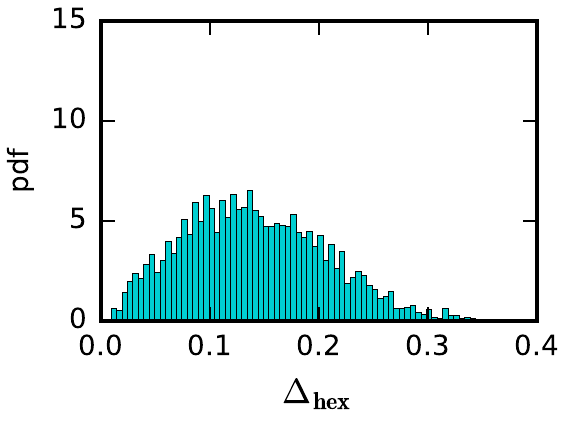}}
	\hspace*{-0.1cm} 
	\subfloat[]{%
	\includegraphics[width=0.24\textwidth]{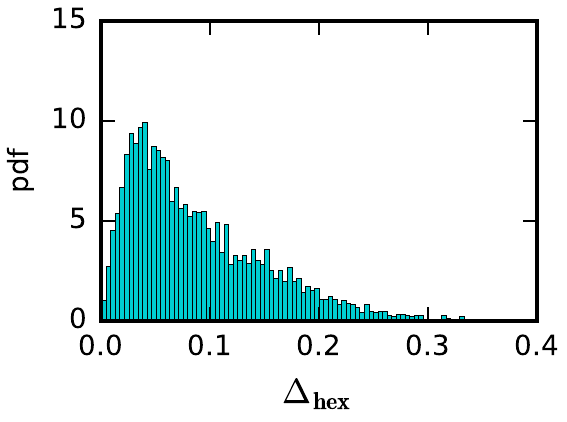}}
	\vspace{-0.5cm} 
	\hspace*{-0.2cm}
	\subfloat[]{%
	\includegraphics[width=0.24\textwidth]{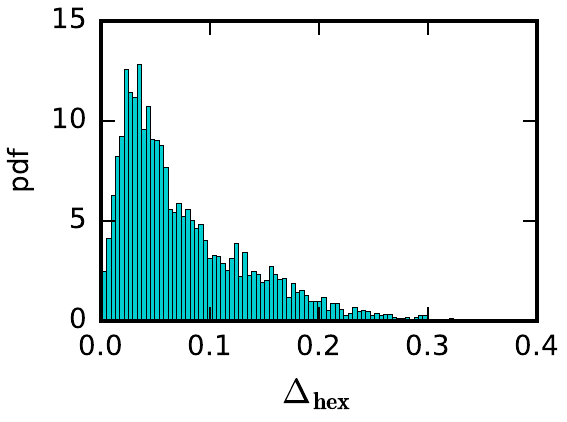}}
	\hspace*{-0.1cm} 
	\subfloat[]{%
	\includegraphics[width=0.24\textwidth]{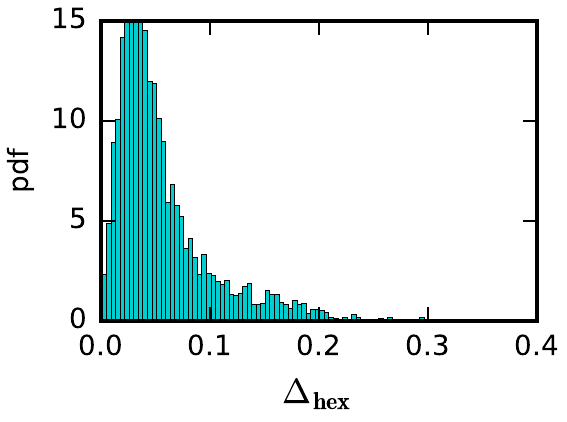}}
\caption{\label{fig-deltahist} \small Typical histograms for the $\mathrm{MT4}$ measure as time $t$ evolves representatively shown for experiment X (see Table \ref{tab-exp}). (a) $t=3.00$ s: Before melting a large peak for small $\Delta_{\mathrm{hex}}$ values is a signature of the crystalline state. (b) $t=3.30$ s: The distribution broadens after melting. (c)-(e) $t=4.50$ s, $t=6.00$ s, $t=7.20$ s: During recrystallization the distribution shifts to smaller $\Delta_{\mathrm{hex}}$ values. (f) $t=12.00$ s: For late times the large peak for small $\Delta_{\mathrm{hex}}$ values is recovered in the recrystallized state.}
\end{figure}

The $\mathrm{MT4}$ analysis carried out in this study is done in analogy to the $\mathrm{MT2}$ analysis. The symmetry metric $ \Delta_{\mathrm{hex}}=\Delta(K_{\mathrm{voronoi} \, \mathrm{cell}},K_{\mathrm{hex}})$ Eq. (\ref{eq-delta}) is calculated locally for every Voronoi cell $K_{\mathrm{voronoi} \, \mathrm{cell}}$ and for every time step in the simulation. $K_{\mathrm{hex}}$ denotes the ideal hexagonal unit cell. In order to distinguish ordered from disordered sections a cut off value of $\Delta_{\mathrm{thresh}} < \Delta_{\mathrm{defects}}$ (corresponding to Voronoi cells close to hexagonal symmetry interpreted as in the crystalline state) is chosen. The specific value that is chosen for $\Delta_{\mathrm{thresh}}$ is indicated in each case in the result section \ref{sec-results}. The number fraction of these particles identified as in the crystalline state will be referred to as $\mathrm{MT4}$ measure in the following. Typical histograms for $\Delta_{\mathrm{hex}}$ for the analyzed recrystallization processes are shown in Fig. \ref{fig-deltahist} as time evolves.
\subsection{Clustering \label{subsec-clustering}}
The area of ordered domains $A_i$ is proportional to the number of particles in the domain $N_d$, weighted with the square of the particle separation $\Delta^2$.  The boundary length $l_i$ is proportional to the number of particles $N_s$ that the boundary line consists of, weighted with the particle separation $\Delta$. It follows that the hypothesis Eq. \ref{eq-fractal} can be reduced to:
\begin{equation}
\langle A_i \rangle \propto \langle l_i \rangle^{1+\alpha}  \label{eq-arealength}
\end{equation}
In order to measure the area $A_i$ and boundary length $l_i$ of the ordered domains, the clustering algorithm DBSCAN (Density-Based Spatial Clustering of Applications with Noise) \cite{dbscan_url} was used. After sorting out the defect particles as identified by the $\Psi_6$ or Minkowski tensor methods, the remaining particles in the crystalline state domains where sorted in clusters via the DBSCAN algorithm. It sorts point clouds into clusters with at least $n_{ \mathrm{min}}$ particles having at most $d_{ \mathrm{max}}$ separation. In order to be able to disjoin clusters linked only by a small number of particles, the DBSCAN algorithm was run two consecutive times with adapted parameter $n_{\mathrm{min}}$: In the second run $n_{\mathrm{min}}$ was increased from $n_{\mathrm{min}}=3$ in the first run (i.e. the smallest clusters have at least four particles), to $n_{\mathrm{min}}=4$. The parameter $d_{\mathrm{max}}$ is set in the range of the mean particle displacement as $d_{\mathrm{max}}=0.75 \mathrm{mm}$ (i.e. the largest particle distance within a cluster can not exceed $0.75 \mathrm{mm}$). Domains in contact with the image boundary were discarded. This restricts the maximum domain size, but for domains not completely within the field of view an estimate of their area and circumference is not possible.

After identifying the particles in domains separated by defect lines and associating them in clusters, as described in the paragraph above, the area and boundary length of each domain could be measured as follows: The concave hull (also known as "alpha shape") of the set of points was calculated as the polygon that represents the area occupied by this set of points in the plane. To achieve this, at first the convex hull and Delaunay triangulation is calculated. The convex hull is comprised of all triangles of the Delaunay triangulation. To get to a concave hull the largest triangles (i.e. the triangles at the boundary of the convex hull) are then discarded from the convex hull: All triangles (with edge lengths $a,b,c$ and area $A$) with radius filter $r_{\mathrm{f}} = abc/(4A) > 1/\gamma$ for an arbitrary parameter $\gamma$. The specific value that is chosen for $1/\gamma$ is indicated in each case in the result section \ref{sec-results}. Only values in the range $1/ \gamma \in \left[0.04, 0.08 \right] \mathrm{mm}$ are considered: For $1 / \gamma > 0.08 \mathrm{mm}$, the algorithm breaks down since domains become internally disconnected until no connected regions can be found anymore. Domains for $1 / \gamma < 0.04 \mathrm{mm}$  are unphysical since the concave hull then includes particles that are not part of the domain as determined by the DBSCAN algorithm.

\begin{figure}[!tbp]
	\captionsetup[subfigure]{position=top,singlelinecheck=off,labelfont=bf,textfont=normalfont, justification=raggedright, margin=10pt,captionskip=-1pt }
	\hspace*{-0.2cm} 
	\subfloat[]{%
	\includegraphics[width=0.24\textwidth]{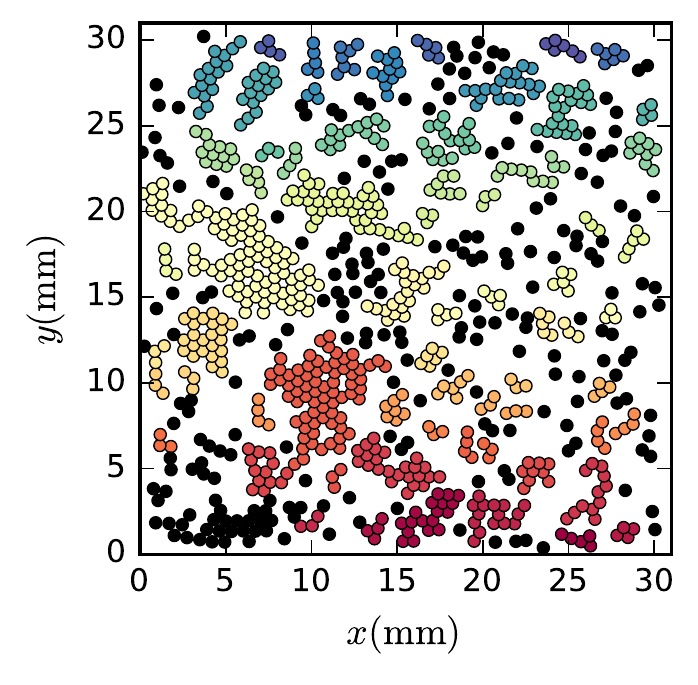}}
	\hspace*{-0.1cm} 
	\subfloat[]{%
	\includegraphics[width=0.24\textwidth]{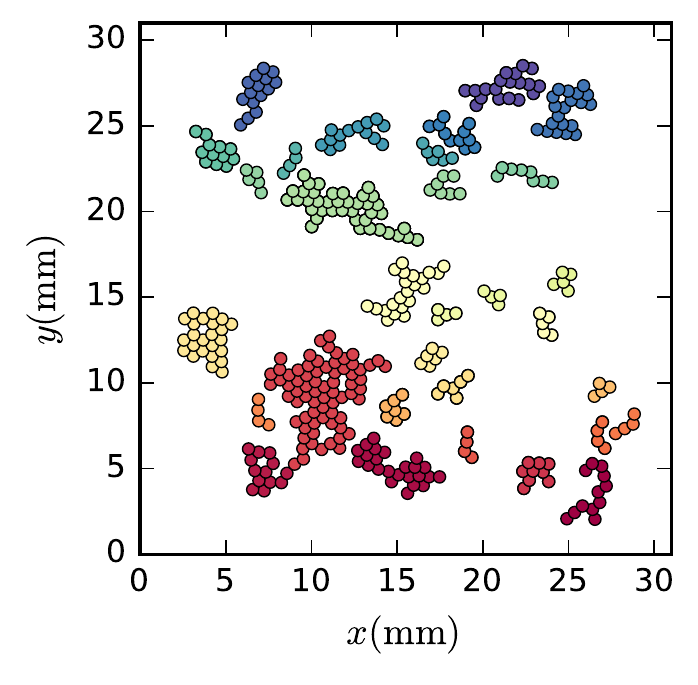}}
	\vspace{-0.5cm} 
	\hspace*{-0.2cm}
	\subfloat[]{%
	\includegraphics[width=0.24\textwidth]{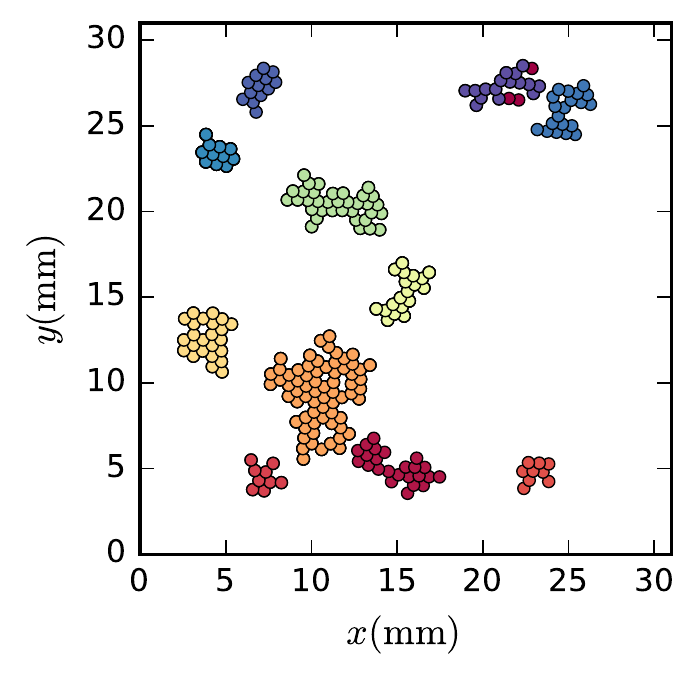}}
	\hspace*{-0.1cm} 
	\subfloat[]{%
	\includegraphics[width=0.24\textwidth]{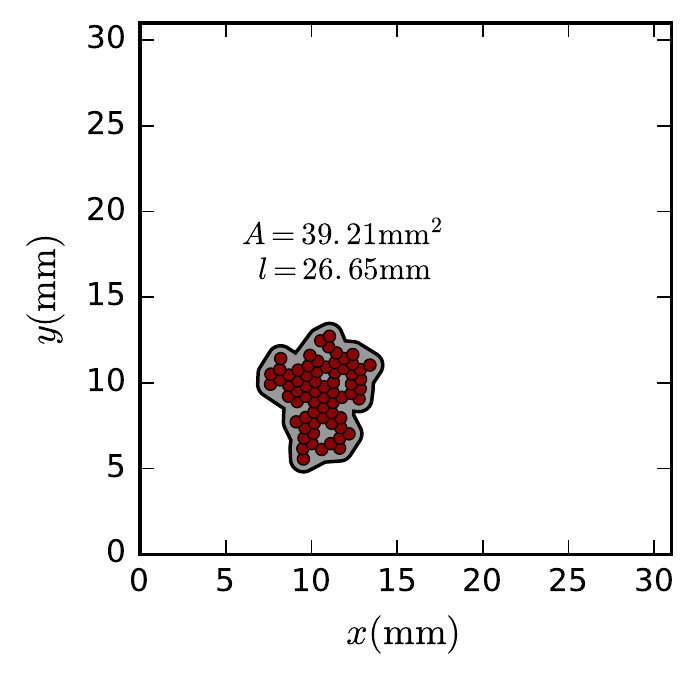}}
\caption{\label{fig-cluster} \small Calculating the area and circumference of crystalline domains: (a) DBSCAN clustering for a specific time-step $t=6.59 \mathrm{s}$. Colors indicate different clusters, black are particles that are not considered as part of a cluster. (b) Clusters touching image boundary and particles not considered in a cluster are deleted. (c) The second DBSCAN clustering removes noise: Very small clusters, small cluster extensions and separates clusters connected by only few particles. (d) Estimate of the area and circumference of a specific domain via a concave hull algorithm. For details consult Sec. \ref{subsec-clustering}.}
\end{figure}

In a last step, the concave hull polygon is smoothed out by buffering it as a smooth contour constructed by discs with a radius $r_{\mathrm{b}}$ in the range of the mean inter-particle separation $r_{\mathrm{b}}=0.75 \mathrm{mm}$. 
An example of the clustering steps is given in Fig. \ref{fig-cluster}.

\subsection{Energy calculation \label{subsec-temp}}
The velocities of every particle are obtained by tracking each particle frame by frame and comparing consecutive images. This provides trajectories in time from which the velocity of every particle is derived. After fitting a Normal distribution to the histogram of velocities in $x$ and $y$ direction separately at every time step, the mean of the width of these histograms gives the particle kinetic energy $E$ (representative of $T$ from Sec. \ref{sec-theory}) for each data set. With this method energies could be resolved down to a level of $0.1 \mathrm{eV}$.
\section{Results \label{sec-results}}

\subsection{Scaling Behavior of the Bond Correlation Function $g_6(r)$\label{subsec-corr}}
\begin{figure}[!tbp]
	\captionsetup[subfigure]{position=top,singlelinecheck=off,labelfont=bf,textfont=normalfont, justification=raggedright, margin=10pt,captionskip=-1pt }
	\vspace*{-0.2cm} 
	\hspace*{-0.2cm} 
	\subfloat[]{%
	\includegraphics[width=0.45\textwidth]{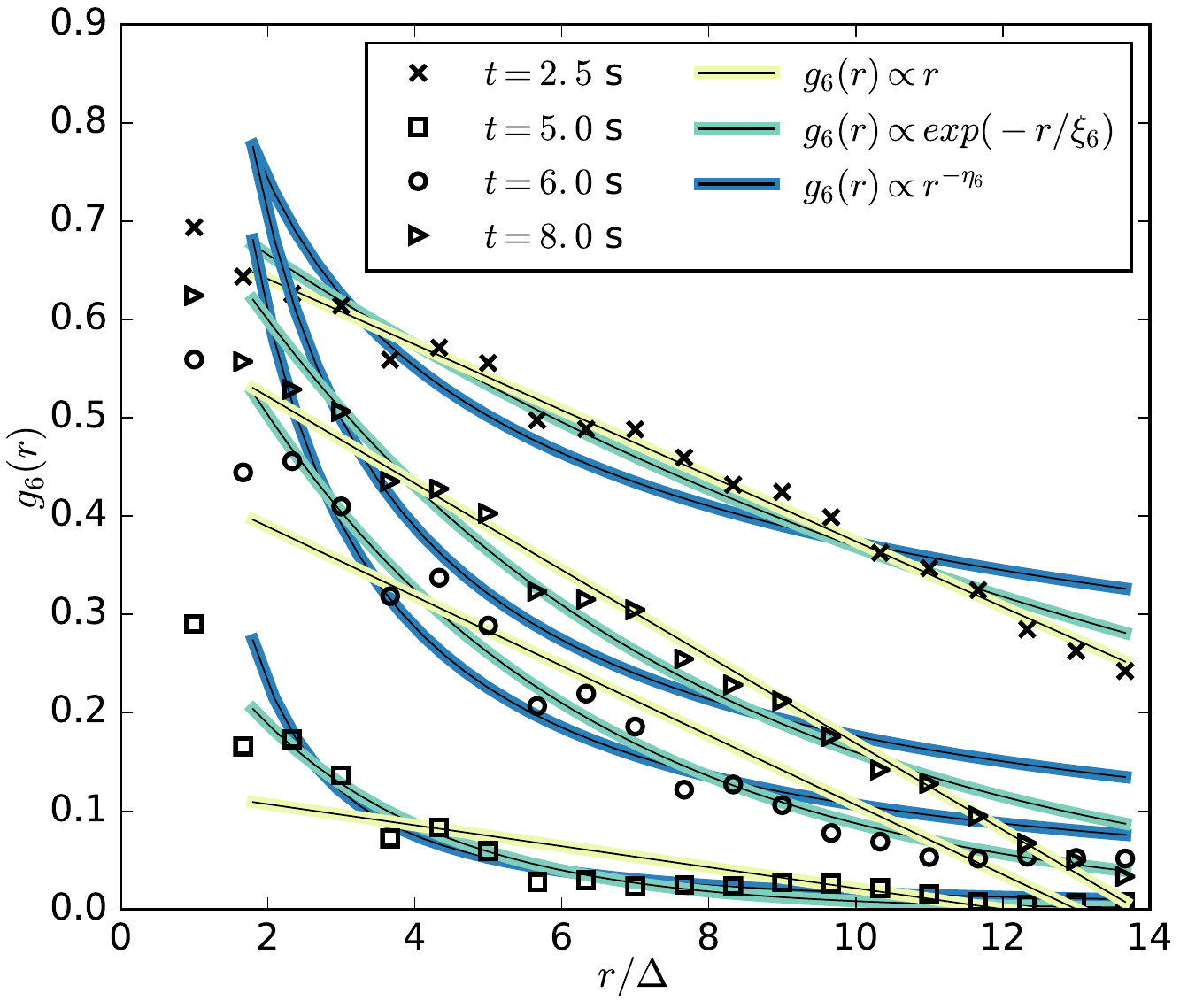}}
	\hspace*{-0.15cm} 
	
	\subfloat[]{%
	\includegraphics[width=0.45\textwidth]{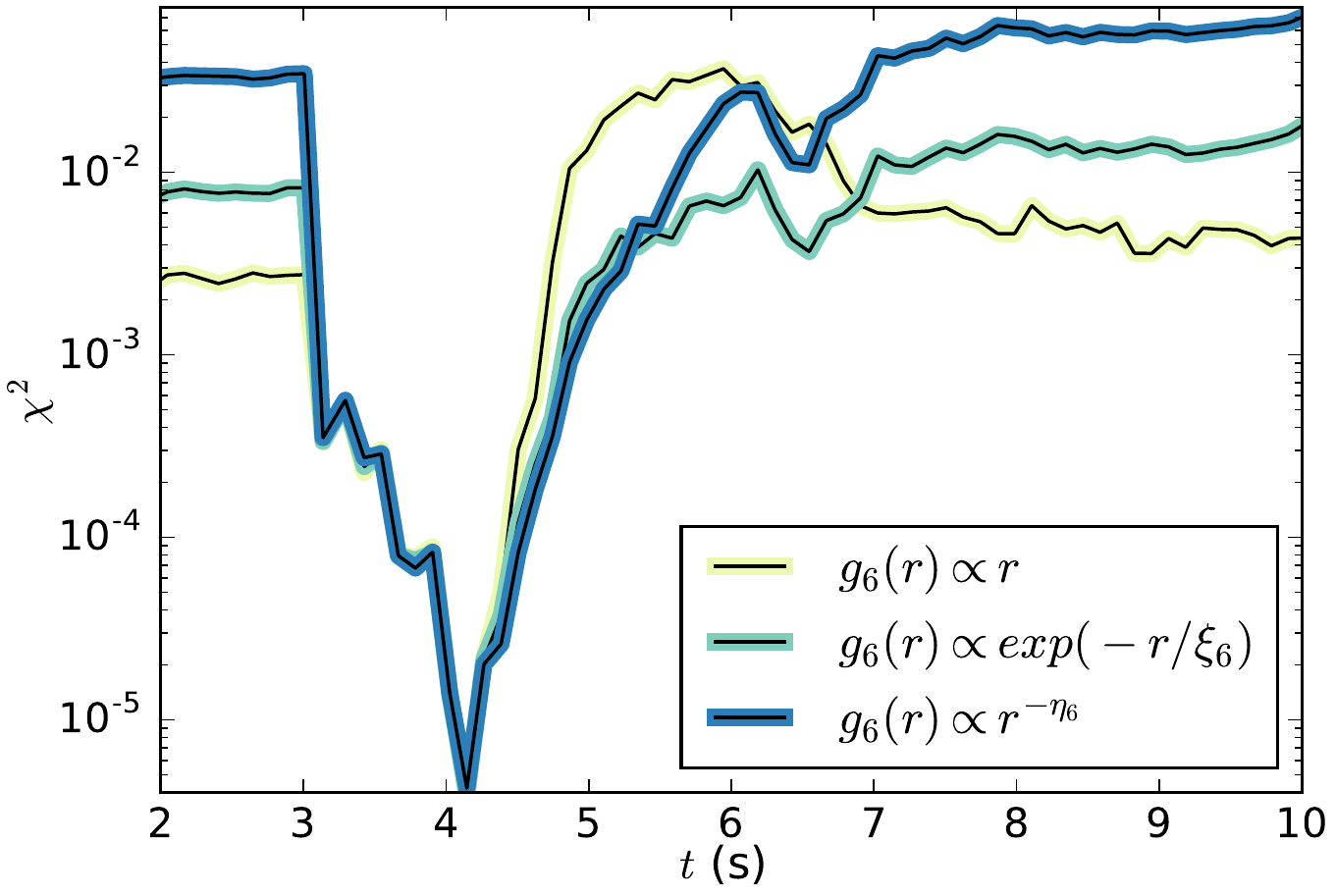}}
	\hspace*{-0.15cm} 
	
	\subfloat[]{%
	\includegraphics[width=0.45\textwidth]{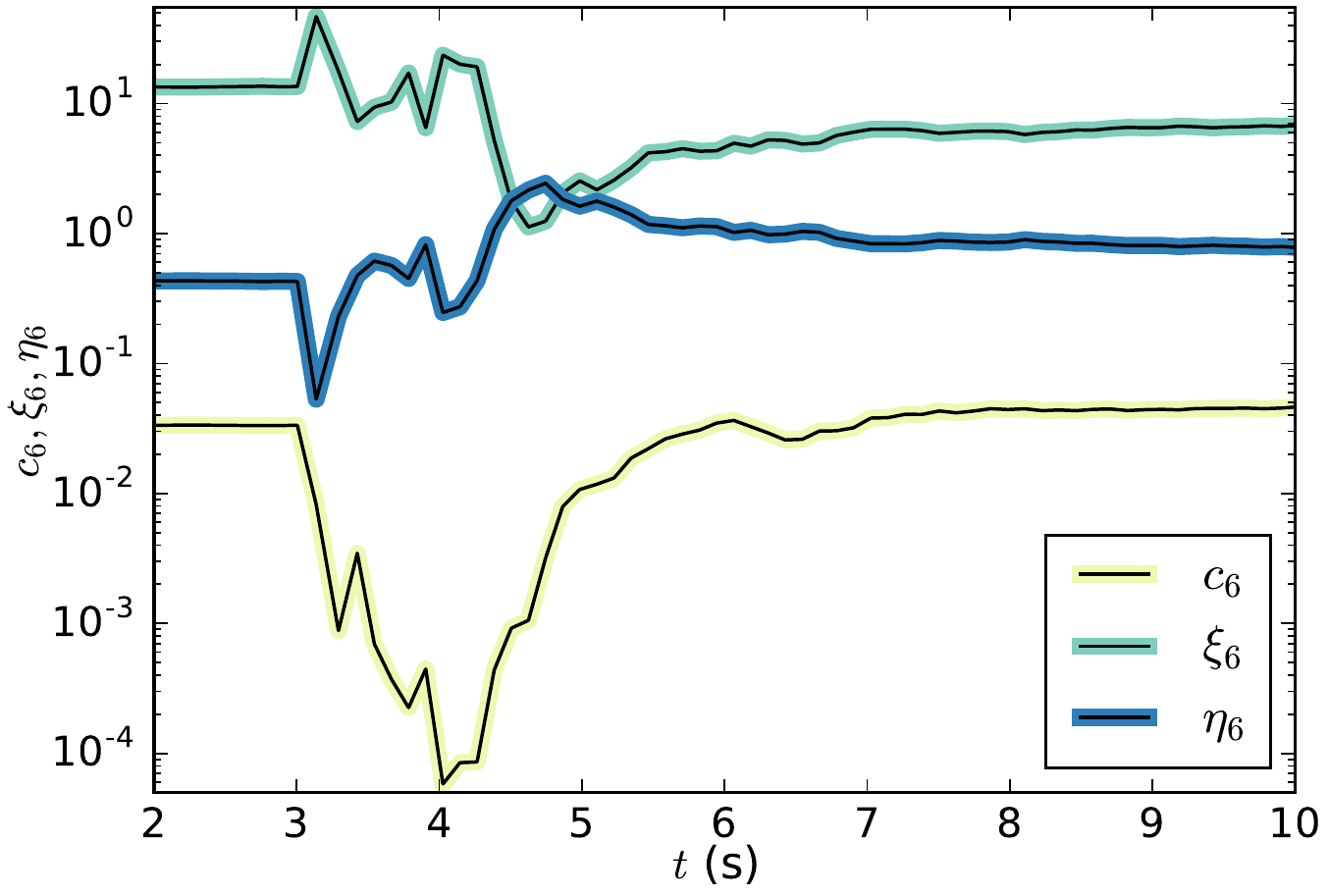}}%
\caption{\label{fig-corr} \small Scaling behavior of the long range decay of the bond correlation function $g_6(r)$. Shown for experimental data set I. (a) Different models are fitted to the long range decay of $g(r)$ at different times $t$. Crystalline state: $g_6(r) \propto c_6 \cdot r$, liquid state: $g_6(r)\propto \mathrm{exp}(-r/\xi_6)$ and hexatic phase: $g_6(r) \propto r^{-\eta_6}$. (b) The chi-squared $\chi^2$ statistic as a measure of the goodness of fit for different decay models. Small values indicate the best model. (c) Values of the best fit parameters for different models. To enhance the clarity of the strongly fluctuating figures during melting times ($\sim 3$ s $<t<5$ s ), panels (b) and (c) only show every 20th data point.}
\end{figure}
The bond correlation function $g_6(r)$ was calculated for all time-steps and datasets and fitted to a linear decay model (crystalline state), an exponential decay model (liquid state) and a power-law decay model (hexatic phase). For brevity of presentation only results of fits for experiment I are shown in Fig. \ref{fig-corr}. The findings, however, are qualitatively the same for all data sets. Panel (a) shows fits of the long-range decay behavior of $g_6(r)$ for different time steps. The goodness of fit $\chi^2$ statistic is shown in Panel (b). Small values indicate high goodness of fit and confidence of the validity of the underlying model. Panel (c) provides the values of the fit parameters $c_6$, $\xi_6$ and $\eta_6$.

For small times, before melting ($\sim 0$ s $<t<3$ s ), and for large times($\sim7$ s $<t<12$ s ), after crystallization, we find the best model to be the linear decay. The linear decay evidences a state of crystalline domains that exhibit internal orientational order but have the freedom to rotate their orientation compared to neighboring domains \cite{15o}.

For times, directly before the linear decay ($\sim 5.5$ s $<t<7$ s) evidences the crystalline state to be the best model, the exponential decay model provides the best goodness of fit (i.e. smallest $\chi^2$ values in Fig. \ref{fig-corr}), indicating a liquid state \cite{04o, 05o}. This already excludes the possibility of a KTHNY type phase transition since between the liquid and crystal state no evidence for the existence of a hexatic phase is found.

In a very short time frame between the chaotic melting and the liquid state ($\sim4.5$ s $<t<5.5$ s) the power-law decay and exponential decay model both provide high goodness of fits values (i.e. low $\chi^2$ values). The power-law decay would indicate a hexatic phase \cite{04o, 05o} in the KTHNY model. However, the KTHNY model predicts a power-law exponent of $\eta < 0.25$ for the hexatic phase. Here, we have much larger values of $\eta > 3$ for all times but the chaotic melting regime where no reliable fit could be performed. Also, the temperature regime does not correspond to a possible hexatic phase. 

Thus, no hexatic state can be found for this phase transition. This implies that the two-dimensional complex plasma phase transitions analyzed in this study are not consistent with the KTHNY theory.

\subsection{Relationship between Domain Area and Boundary Length \label{subsec-frac-area}}
\begin{figure}[!tbp]
	\captionsetup[subfigure]{position=top,singlelinecheck=off,labelfont=bf,textfont=normalfont, justification=raggedright, margin=10pt,captionskip=-1pt }
	\vspace*{-0.2cm} 
	\hspace*{-0.2cm} 
	\subfloat[]{%
	\includegraphics[width=0.45\textwidth]{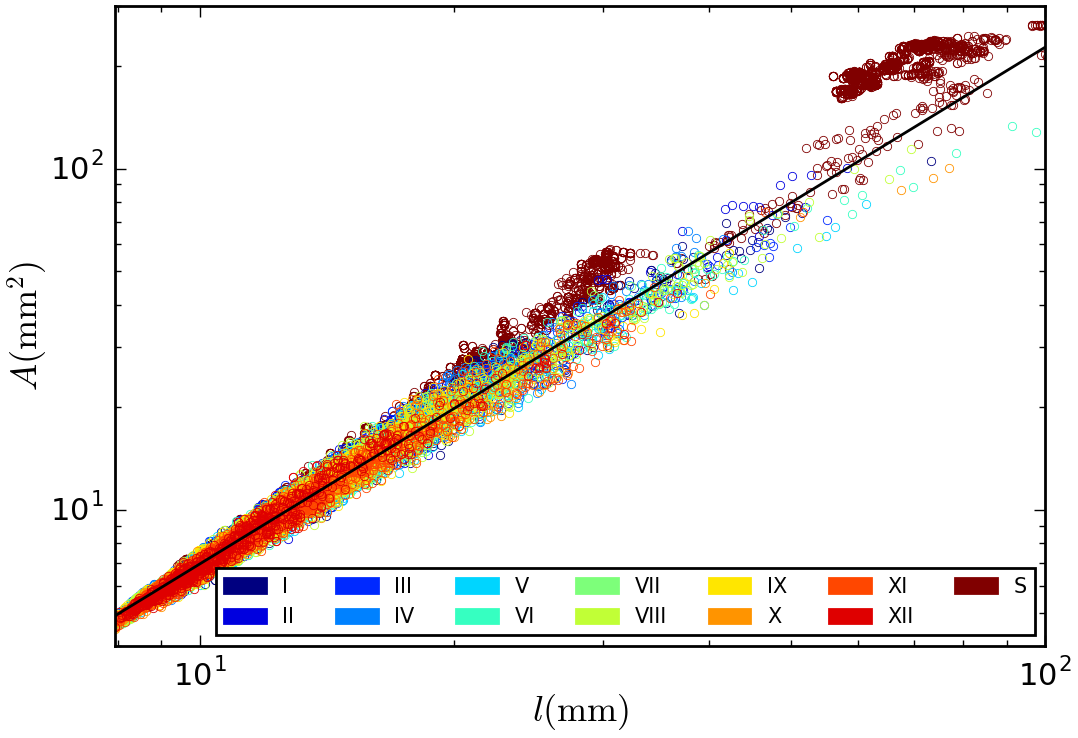}}
	\hspace*{-0.15cm} 
	
	\subfloat[]{%
	\includegraphics[width=0.45\textwidth]{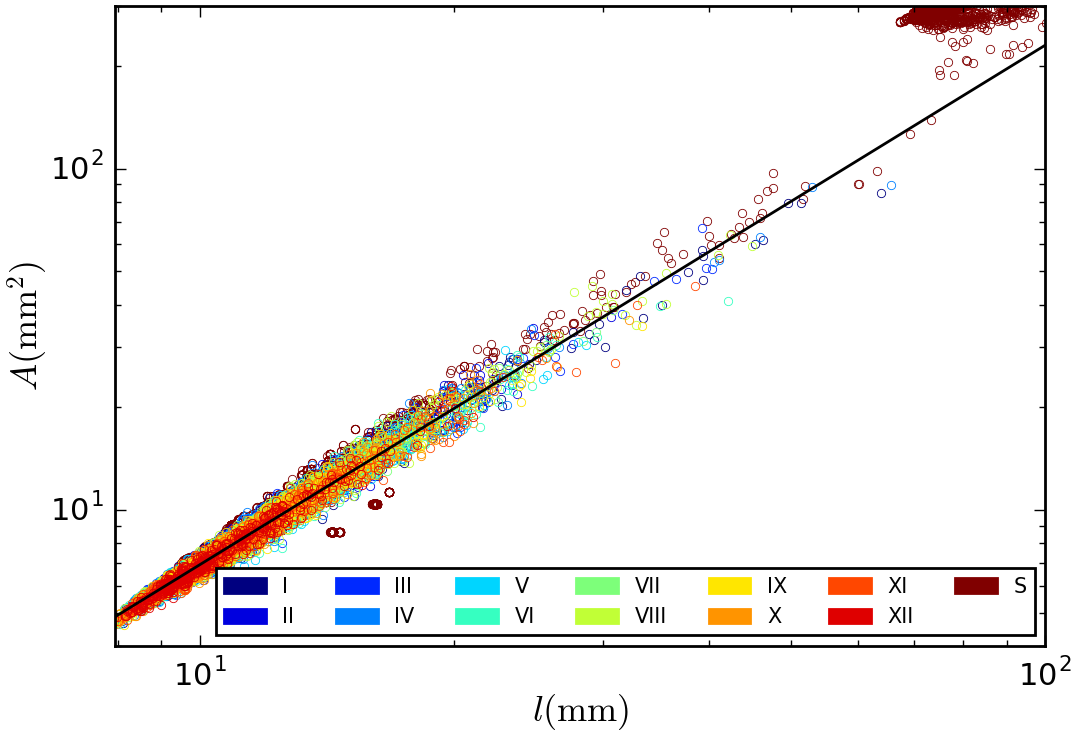}}
	\hspace*{-0.15cm} 
	
	\subfloat[]{%
	\includegraphics[width=0.45\textwidth]{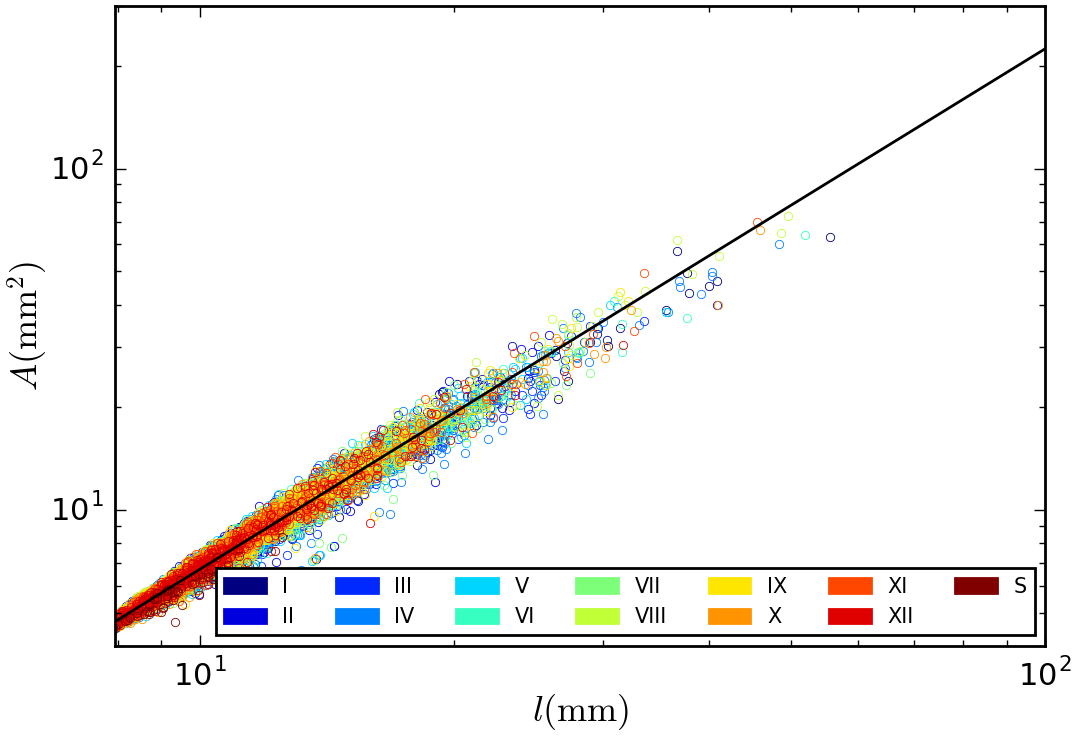}}%
\caption{\label{fig-area_length} \small The area $A_{i}$ of crystalline domains plotted against their boundary length $l_i$. Different colors indicate different experiments and the simulation. The solid line is the mean of all least square linear fits to the power law Eq. \ref{eq-arealength} $\langle A_i \rangle \propto \langle l_i \rangle^{1+\alpha}$ for the experiments and the simulation. In Fig. (a) defects are identified via a $\Psi_6$ bond order metric (Sec. \ref{subsec-psi6},$\Psi_{6,\mathrm{thresh}}=0.5$ ), in Fig. (b) the particles in crystalline states are identified via the $\mathrm{MT2}$ isotropy index method (Sec. \ref{subsec-beta}, $\beta_{\mathrm{thresh}}=0.81$) and in Fig. (c) they are identified via the $\mathrm{MT4}$ symmetry metric method as explained in Sec. \ref{subsec-delta} ($\Delta_{\mathrm{thresh}}=0.18$).  Area and boundary length are measured using a DBSCAN clustering algorithm as explained in Sec. \ref{subsec-clustering}. Individual exponent values can be found in Table \ref{tab-arealength}. }
\end{figure}
The results of the analysis with the above explained methods and measures are presented in the following.
In this section the hypothesis introduced in Eq. \ref{eq-fractal} and condensed to Eq. \ref{eq-arealength} is tested via plotting measured domain areas $A_i$ against their circumferences $l_i$ and thus possibly obtaining the power law exponent $\alpha$. To this end, first the defect Voronoi cells were detected via the $\Psi_6$ bond order parameter (Sec. \ref{subsec-psi6}), the $\mathrm{MT2}$ (Sec. \ref{subsec-beta}) and the $\mathrm{MT4}$ (Sec. \ref{subsec-delta}) method. Discarding these defects leaves ordered disjoint domains that are clustered using a DBSCAN clustering algorithm (\ref{subsec-clustering}).
The relation between the area of crystalline domains $A_{i}$ and the boundary length $l_{i}$ of defect lines separating the crystalline domains is shown in Fig. \ref{fig-area_length}. All defect detection methods provide consistent results and every experiment is consistent with the scaling relation $\langle N_d \Delta^2 \rangle \propto \left[ \Delta \langle N_s \rangle \right]^{1+\alpha}$. The exponents for the experiments are consistent with those found for the simulation data. The values of $\alpha$ obtained by least-square fits are listed in Table \ref{tab-arealength} and are consistent with the findings in Sec. \ref{subsec-frac-en}.

In the simulation data we find small deviations. Since they are generated using a parabolic potential the particle density decreases in the radial direction. Thus, the density-based DBSCAN algorithm cannot as easily be applied to the simulation data as in the experimental case. For the $\Psi_{\mathrm{6}}$ and $\mathrm{MT2}$ metric in most cases a large cluster in the center is detected. For the $\mathrm{MT4}$ metric predominantly very small clusters are detected.

\begin{figure}[!tbp]
	\captionsetup[subfigure]{position=top,singlelinecheck=off,labelfont=bf,textfont=normalfont, justification=raggedright, margin=10pt,captionskip=-1pt }
	\vspace*{-0.2cm} 
	\hspace*{-0.2cm} 
	\subfloat[]{%
	\includegraphics[width=0.45\textwidth]{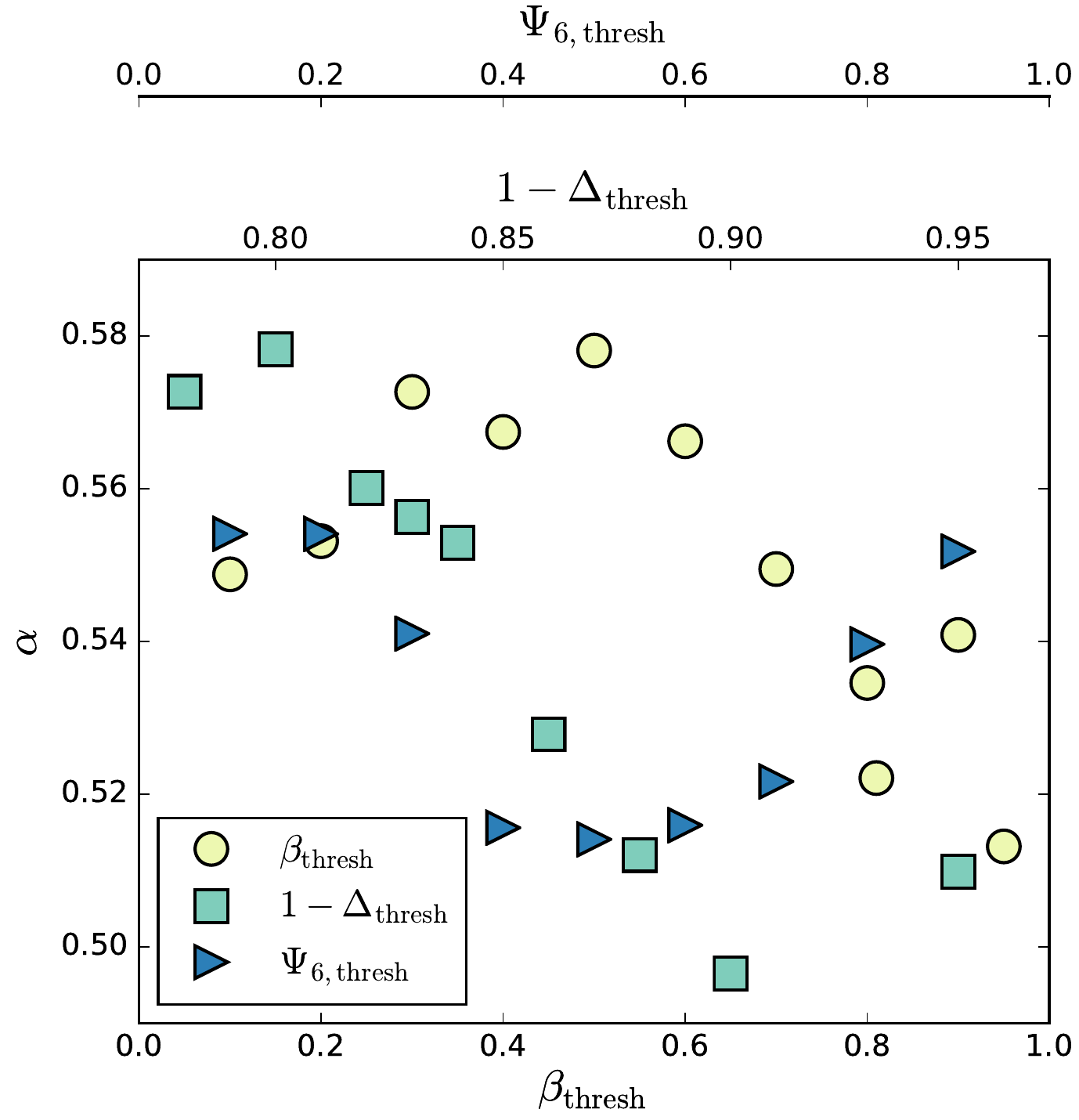}}
	\hspace*{-0.15cm} 
	
	\subfloat[]{%
	\includegraphics[width=0.45\textwidth]{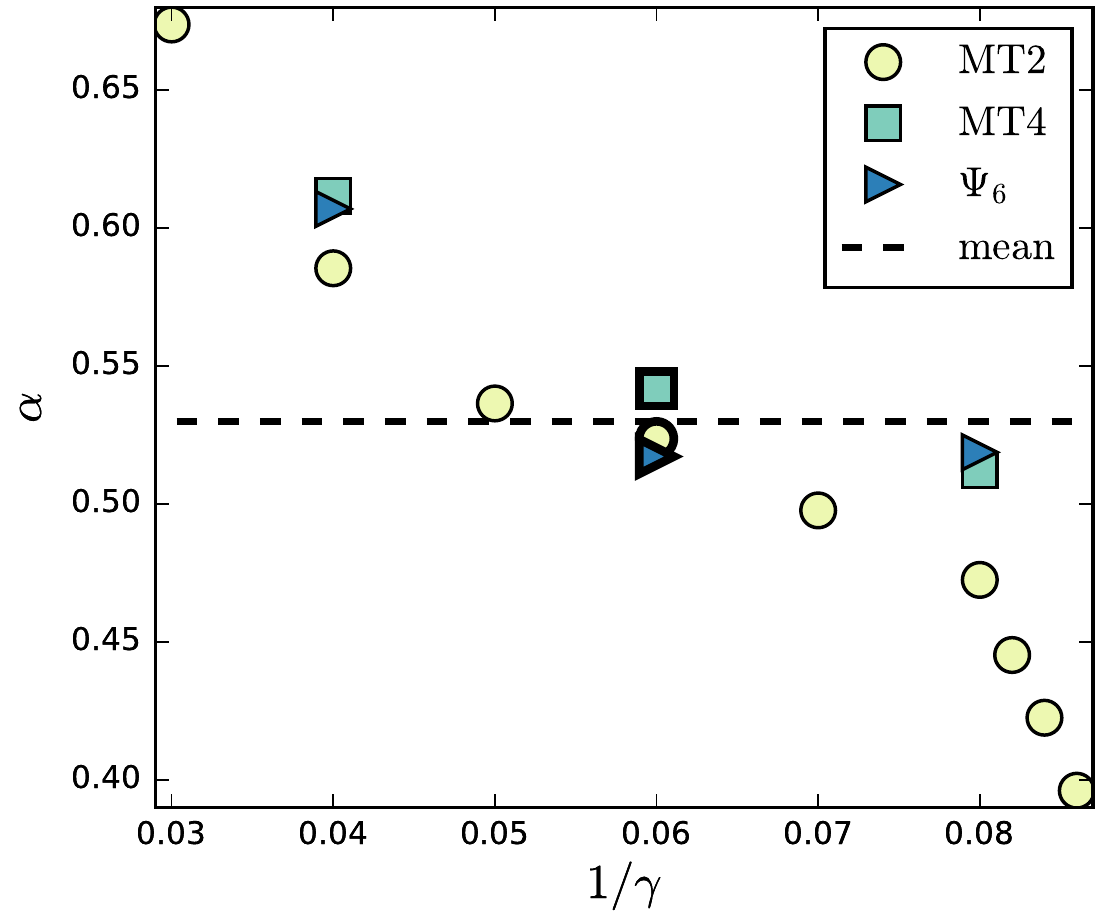}}%
\caption{\label{fig-area_length_univ} \small (a) The exponents $\alpha$ from Eq. \ref{eq-arealength} $\langle A_i \rangle \propto \langle l_i \rangle^{1+\alpha}$, obtained by linear fits of the fractal relationship of domain area and circumference are plotted as the cut-off parameter of the particular measure is varied. $\alpha$ is found to be very stable: Varying the cutoff parameters in their whole range only gives rises to very small changes in $\alpha$. We find variations of $  \sqrt{Var[\alpha_{\beta}]} / \left< \alpha_{\beta} \right> = 2.7 \, \% $, $ \sqrt{Var[\alpha_{\Delta}]} /\left< \alpha_{\Delta} \right>  = 4.3 \, \% $ and $ \sqrt{Var[\alpha_{\Psi_6}]} /\left< \alpha_{\Psi_6} \right>  = 1.9 \, \% $. The DBSCAN parameters are constant, in particular $1/ \gamma =0.06 \, \mathrm{mm}$.
(b) variation of $\alpha$ from Eq. \ref{eq-arealength} $\langle A_i \rangle \propto \langle l_i \rangle^{1+\alpha}$ by variation of the DBSCAN parameter $\gamma$. (The cut-off parameter $\beta_{\mathrm{thresh}}=0.81$ is constant.) The data points with thick marker-edges ($1 / \gamma = 0.06 \, \mathrm{mm}$) are the mean values of the individual $\alpha$ in Table \ref{tab-arealength}. The final mean value is indicated by a dashed line.}
\end{figure}

It is noteworthy that the exponents are very stable (see Fig. \ref{fig-area_length_univ}): They are independent of the cut-off value used in the Minkowski methods to define crystalline cells and independent of the particular choice of Minkowski tensor metric and tensor rank. Even the $\Psi_6$ bond order parameter gives a consistent result. Also, they depend only very weakly on the  particular choice of parameters (reasonable values are discussed below and in Fig. \ref{fig-area_length_univ} (b)) in the DBSCAN clustering algorithm and the particular parameters in the calculation of the domain area and length via the convex hull algorithm. Varying the cutoff parameters in a large range only gives rises to very small changes in $\alpha$. Variations are in the range of only a few percent and are listed in the caption of Fig. \ref{fig-area_length_univ} (a). In this figure the mean values $\langle \alpha \rangle$ are plotted for the whole range of cutoff values for the $\mathrm{MT2}$, $\mathrm{MT4}$ and $\Psi_6$ bond order method respectively.
Changing the smoothing parameter $r_b$ over one order of magnitude has practically no effect.
Varying the DBSCAN parameter $1/ \gamma$ (it can be thought of as describing the raggedness of the concave hull) also gives only small changes in $\alpha$ as depicted in Fig. \ref{fig-area_length_univ} (b). The largest deviations are observed for values $1/ \gamma < 0.04 \mathrm{mm}$ and $1/ \gamma > 0.08 \mathrm{mm}$. Yet, only values of $1/ \gamma \in \left[0.04 ,0.08\right] \mathrm{mm}$ are physically relevant since for too large values the domains become disconnected and for too small values the domains become more and more convex and include neighboring particles that are not part of the detected ordered domains. 
Averaging over all experiments, all defect detection methods with all threshold values and all physical DBSCAN parameters leads to a final value for the scaling exponent:
\begin{equation}
\alpha = 0.52 \pm 0.05 \label{eq-alphaval}
\end{equation} 
This is the mean value (and uncertainty as standard deviation) of measurements as depicted in Fig. \ref{fig-area_length_univ} (b) for the range of physical values $1/ \gamma \in \left[0.04 ,0.08\right] \mathrm{mm}$.

\begin{figure}[!tbp]
	\captionsetup[subfigure]{position=top,singlelinecheck=off,labelfont=bf,textfont=normalfont, justification=raggedright, margin=10pt,captionskip=-1pt }
	\vspace*{-0.2cm} 
	\hspace*{-0.2cm} 
	\subfloat[]{%
	\includegraphics[width=0.32\textwidth]{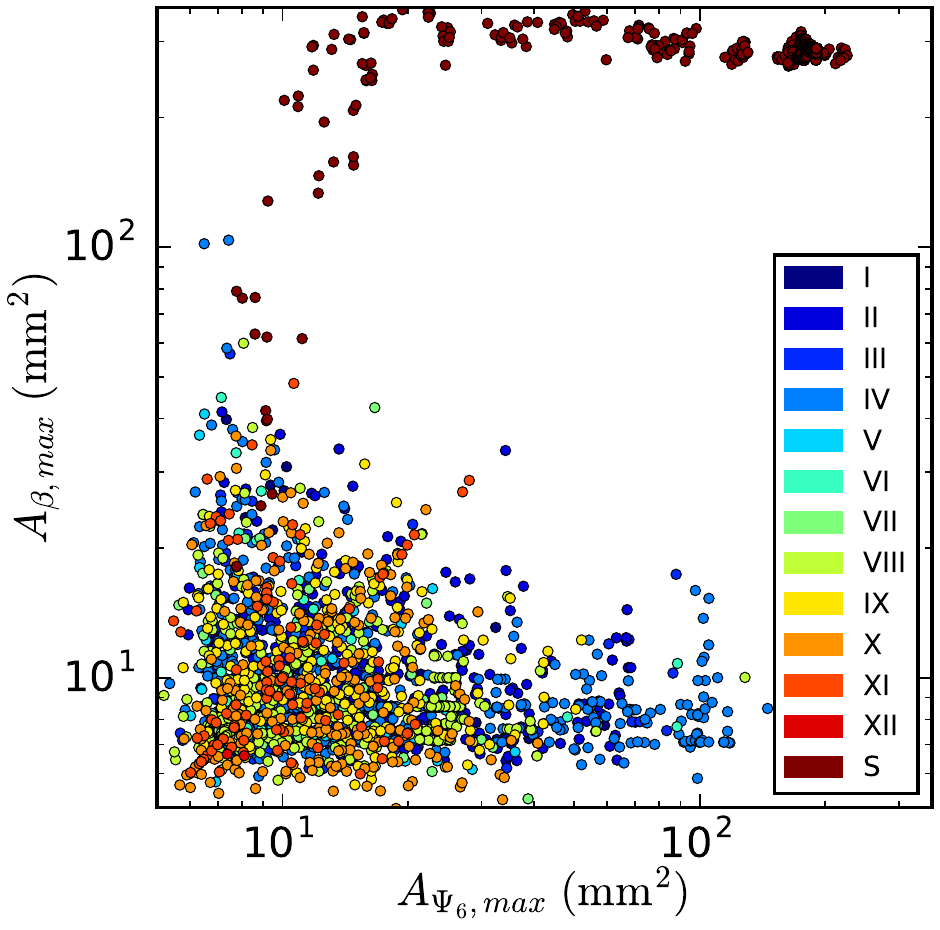}}
	\hspace*{-0.15cm} 
	
	\subfloat[]{%
	\includegraphics[width=0.32\textwidth]{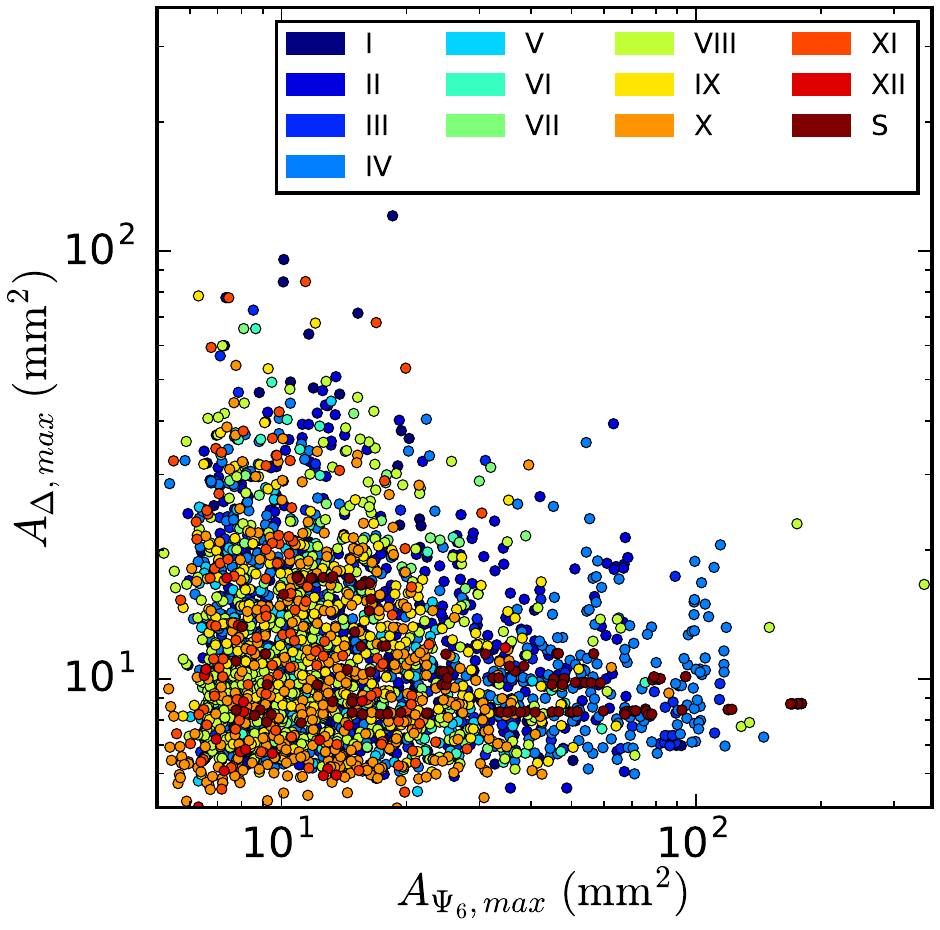}}
	\hspace*{-0.15cm} 
	
	\subfloat[]{%
	\includegraphics[width=0.32\textwidth]{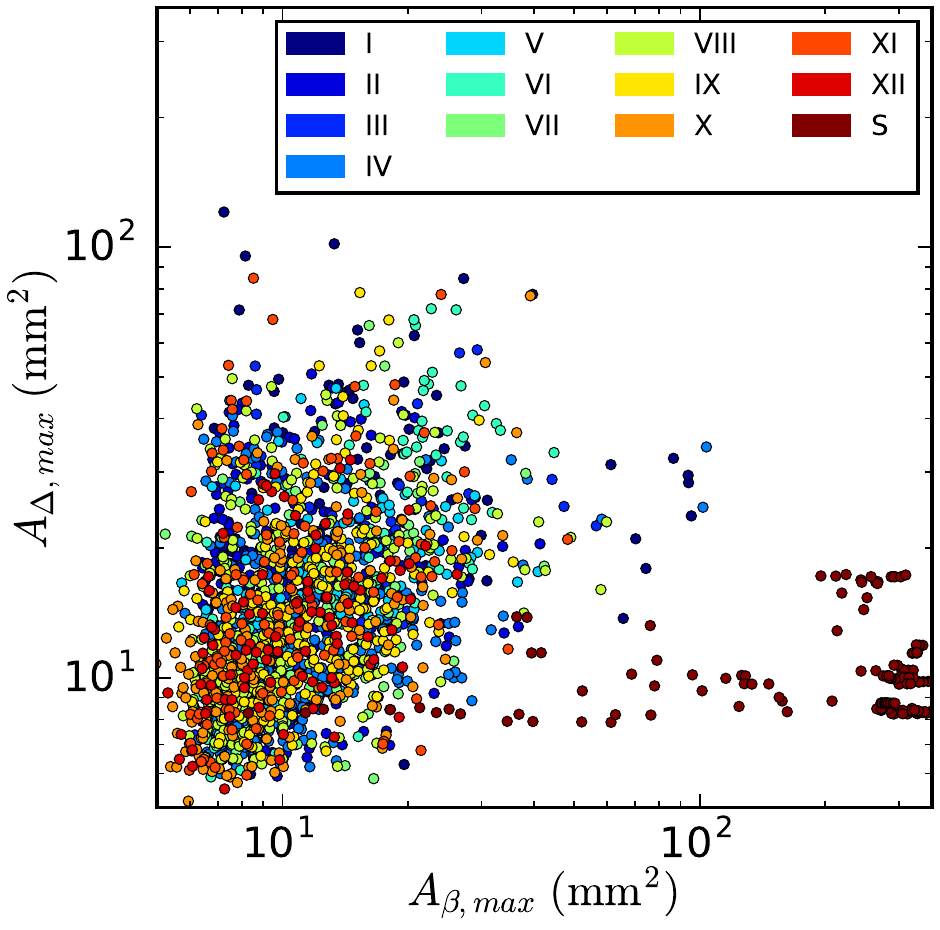}}%
\caption{\label{fig-compare_a} \small A comparison of the detected domain areas $A_i$. For every point in time in which two measures both detect at least one crystalline domain the correlation between the largest domain of each measure is plotted. Comparison between the largest domains for (a) the $\mathrm{MT2}$ and $\Psi_6$ measure, (b) the $\mathrm{MT4}$ and $\Psi_6$ measure, and (c) the $\mathrm{MT2}$ and $\mathrm{MT4}$ measure. The Minkowski measures show the largest correlation, however, the data points are still widely spread. In all graphs (a)-(c) we find the simulation data points to be outliers. This is due to the fact that the clustering algorithm is not applicable in a straight forward way to this data, as discussed in the text.}
\end{figure}

Detected domain sizes for different methods are compared in Fig. \ref{fig-compare_a}). Since multiple domains are detected during one point in time the comparison is only between the largest domain of each measure and point in time. With this method we only find few clusters that are detected consistently using different methods. Nevertheless one can see that the Minkowski measures are more similar to each other than to the $\Psi_6$ bond order metric whereas the higher rank Minkowski measure provides even less similar domains compared to the $\Psi_6$ measure than the lower ranked Minkowski tensor measure.

\begin{table}
\caption{\label{tab-arealength} Power law exponent $\alpha$ for the area-length scaling in Fig. \ref{fig-area_length} measured via the  $\Psi_6$ (\ref{subsec-psi6}, $\Psi_{6,\mathrm{thresh}}=0.7$), the $\mathrm{MT2}$ (\ref{subsec-beta}, $\beta_{\mathrm{thresh}}=0.88$) and the $\mathrm{MT4}$ (\ref{subsec-delta}, $\Delta_{\mathrm{thresh}}=0.12$) methods in Eq. \ref{eq-arealength} $\langle A_i \rangle \propto \langle l_i \rangle^{1+\alpha}$, for experiments I-XII and the simulation (\ref{sec-exp-sim}). For the corresponding graphs consult Fig. \ref{fig-area_length}. The last row is the mean value of all above with the standard deviation as uncertainty. Area and circumference were measured via a DBSCAN algorithm (\ref{subsec-clustering}, here $1/ \gamma = 0.06 \, \mathrm{mm}$, corresponding to the points in Fig. \ref{fig-area_length_univ}(b) marked with thick marker-edges.)}
\begin{ruledtabular}
\begin{tabular}{ccccccc}
 &$\alpha$ ($\Psi_6$)&$\alpha$ (MT2)&$\alpha$ (MT4) \\
\hline
\rm I & 0.526 & 0.505 & 0.519 \\
\rm II & 0.569  & 0.550 & 0.570  \\
\rm III & 0.503  & 0.535 & 0.539  \\
\rm IV & 0.550  & 0.524 & 0.566  \\
\rm V & 0.466  & 0.487 & 0.574  \\
\rm VI & 0.485  & 0.494 & 0.547  \\
\rm VII & 0.515  & 0.540 & 0.558  \\
\rm VIII & 0.501  & 0.545 & 0.547  \\
\rm IX & 0.527  & 0.518 & 0.570 \\
\rm X & 0.488  & 0.545 & 0.545 \\
\rm XI & 0.490  & 0.479 & 0.578 \\
\rm XII & 0.548  & 0.543 & 0.560 \\
\rm S & 0.752 & 0.856 & 0.529 \\
\hline
$\langle$ \rm I...XII $\rangle$ & 0.51 $\pm$ 0.02 & 0.52 $\pm$0.02 & 0.55 $\pm$ 0.02 \\
\end{tabular}
\end{ruledtabular}
\end{table}

\subsection{Relationship between Energy and Defect Fraction \label{subsec-frac-en}}

\begin{figure}[!tbp]
\centering
	\captionsetup[subfigure]{position=top,singlelinecheck=off,labelfont=bf,textfont=normalfont, justification=raggedright, margin=10pt,captionskip=-1pt }
	\vspace*{-0.2cm} 
	\hspace*{-0.2cm} 
	\subfloat[]{%
	\includegraphics[width=0.45\textwidth]{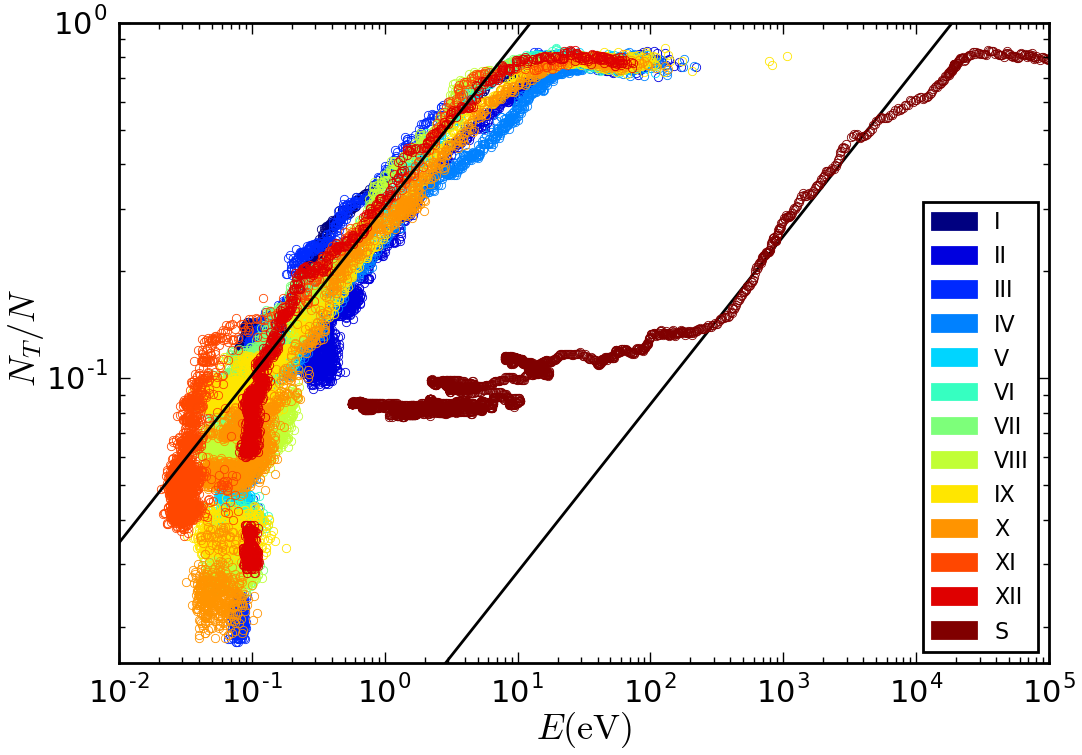}}
	\hspace*{-0.15cm} 
	
	\subfloat[]{%
	\includegraphics[width=0.45\textwidth]{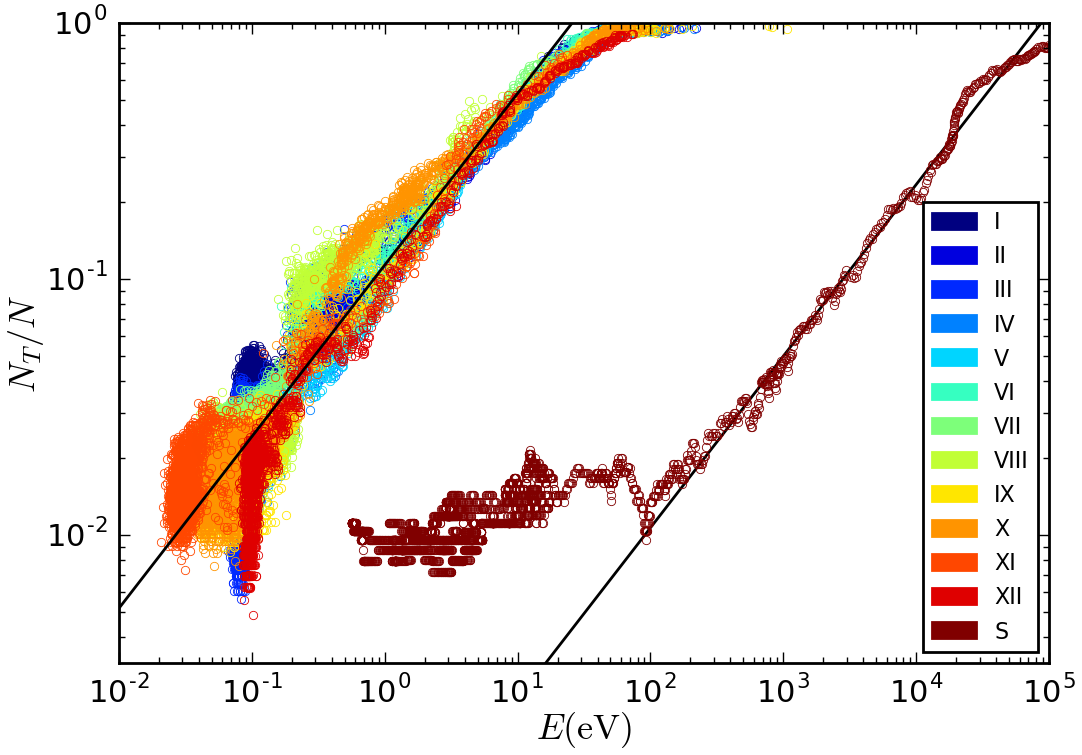}}
	\hspace*{-0.15cm} 
	
	\subfloat[]{%
	\includegraphics[width=0.45\textwidth]{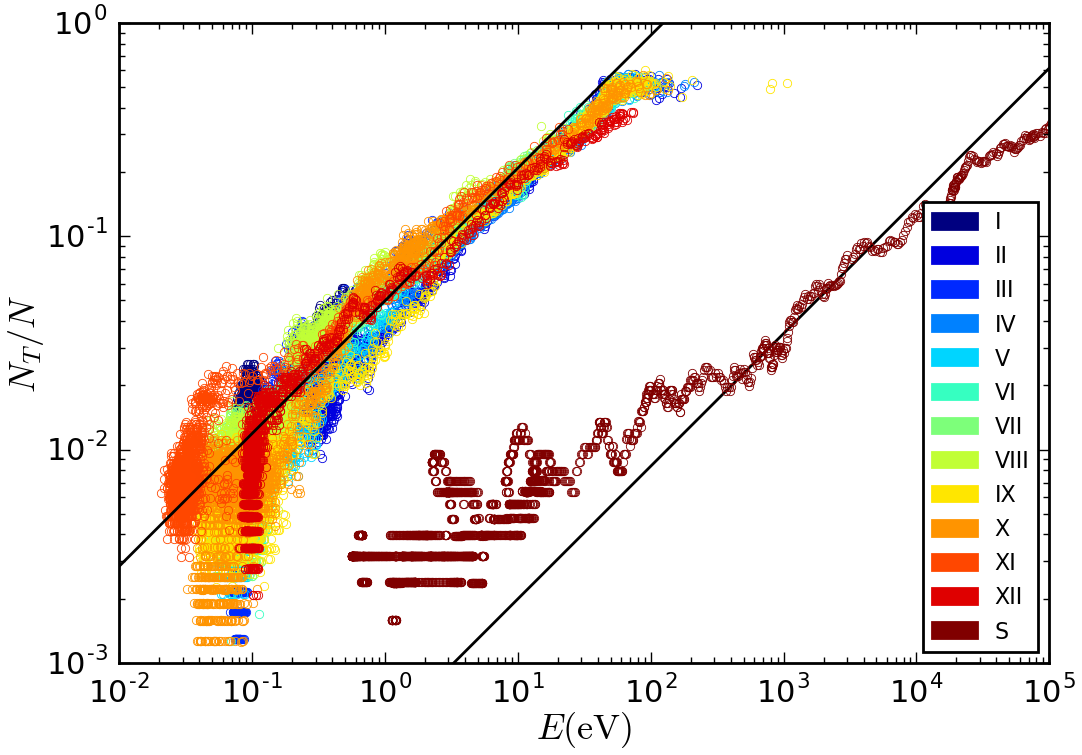}}%
\caption{\label{fig_defect_energy} \small The defect fraction $N_{T}/N$ plotted against the particle kinetic energy $E$. Different colors indicate different experiments and the simulation. The solid line is the mean of all linear fits to the power law $N_T / N \propto E^{\xi}$ for all experiments. The linear fit of the experimental data is shifted onto the simulation data in a parallel fashion. In (a) the defect fraction is obtained via the $\Psi_6$ method ($\Psi_{6,\mathrm{thresh}}=0.5$). The exponent is $ \alpha_{\Psi6}=0.313$. The fit is constrained to energies in the interval $E \in \left[10^{-1}, 10^{1} \right] \mathrm{eV}$. (b) MT2 ($\beta_{\mathrm{thresh}}=0.81$); $ \alpha_{\mathrm{MT2}}=0.518$; fit interval $E \in \left[10^{-1}, 10^{1.7} \right] \mathrm{eV} $. (c) MT4 method ($\Delta_{\mathrm{thresh}}=0.18$); $ \alpha_{\mathrm{MT4}}=0.507$; fit interval $E \in \left[10^{-1}, 10^{2} \right] \mathrm{eV} $. Individual exponent values can be found in Table \ref{tab-defect}.}
\end{figure}

In this section the theoretical prediction of Eq. \ref{eq-defectenergy} $N_T / N \propto T^{2 \alpha/\left( 1+\alpha \right)}\propto E^{2 \alpha/\left( 1+\alpha \right)}$ of the FDS theory \cite{12o} is tested via plotting defect fractions $N_T/N$ against the kinetic energy $E$ of the particles. This is shown in Fig. \ref{fig_defect_energy}. One can clearly see that the relation can well be described by a power law within a reasonable energy interval. After the detection of defects their number fraction $N_T / N$ is obtained by simple division of the total defect number $N_T$ and the total particle number $N$. The system temperature $E=k_B T$ is determined by fitting a normal distribution to the histogram of each component of the particle velocity vectors. Plotting these values in a log-log plot (Fig. \ref{fig_defect_energy}) then gives the power-law exponent $\alpha$ for every data set and method. The exponents obtained in subsections Sec. \ref{subsec-frac-area} and Sec. \ref{subsec-frac-en} are compared.

\begin{figure}[!tbp]
\centering
	\captionsetup[subfigure]{position=top,singlelinecheck=off,labelfont=bf,textfont=normalfont, justification=raggedright, margin=10pt,captionskip=-1pt }
	\vspace*{-0.2cm} 
	\hspace*{-0.2cm} 
	\subfloat[]{%
	\includegraphics[width=0.32\textwidth]{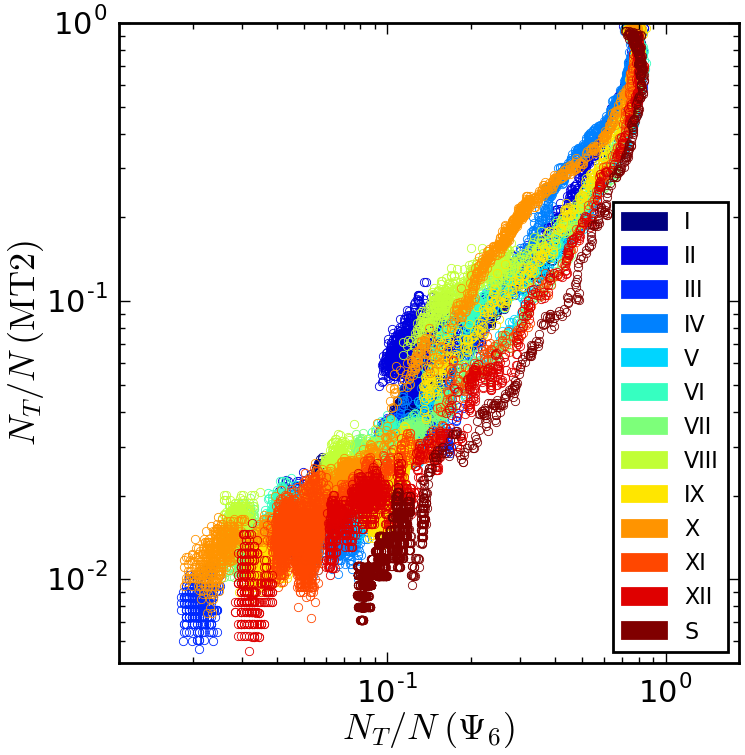}}
	\hspace*{-0.15cm} 
	
	\subfloat[]{%
	\includegraphics[width=0.32\textwidth]{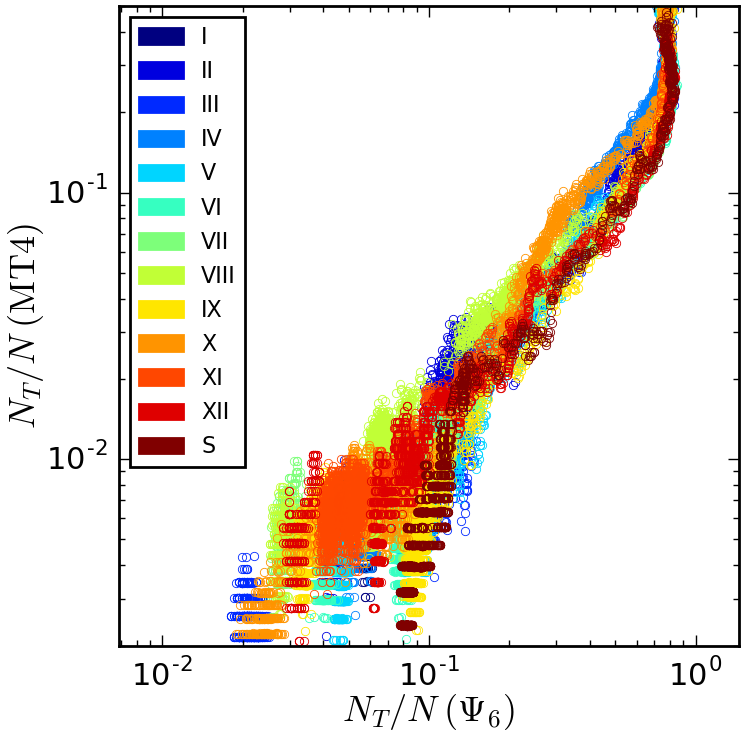}}
	\hspace*{-0.15cm} 
	
	\subfloat[]{%
	\includegraphics[width=0.32\textwidth]{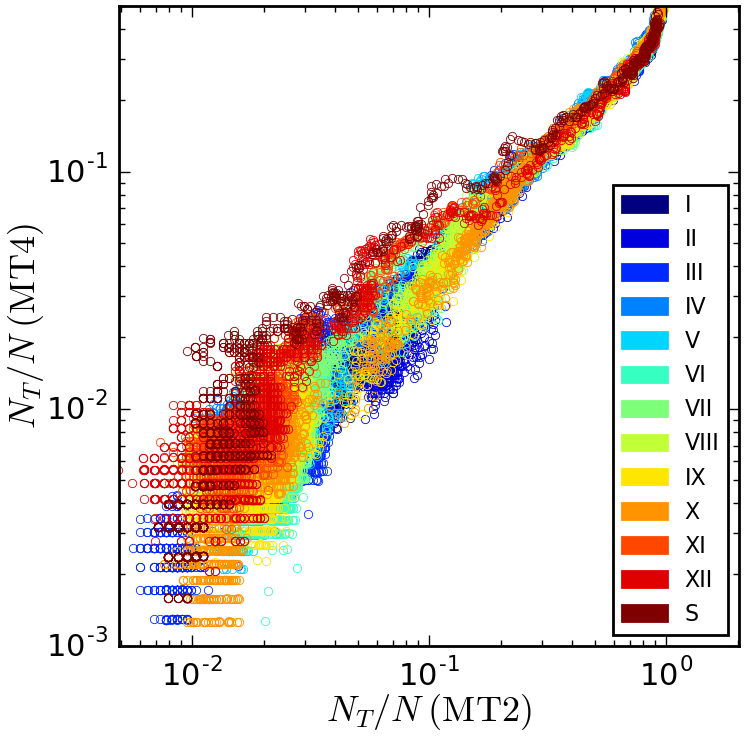}}%
\caption{\label{fig-defect_compare} \small  A comparison of the detected defect fraction $N_T/N$ at equal energies $E$. (a) $\mathrm{MT2}$ vs. $\Psi_6$, (b) $\mathrm{MT4}$ vs. $\Psi_6$, (c) $\mathrm{MT2}$ vs. $\mathrm{MT4}$. The Minkowski measures resolve defect fractions for energy levels about one order of magnitude larger than the bond order metric. The higher rank Minkowski tensor measure resolves defect fractions even further than the lower ranked tensor measure. Compared to the Minkowski tensor measures the bond order metric shows an early saturation for high energies.}
\end{figure}	

The dependence of the defect fraction $N_{T}/N$ on the kinetic energy is shown in Fig. \ref{fig_defect_energy}. Every measure and experiment is consistent with a power law Eq. \ref{eq-defectenergy} $N_{T}/N \propto E^{\xi}$, with $\xi=2 \alpha / (1+ \alpha)$. The exponents obtained by least-square fits are listed in Table \ref{tab-defect}. The exponents for the experiments are comparable with the exponents found for the simulation data: The line of best fit, shifted in a parallel fashion to simulation data energies, fits the data quite well. Thus, the simulation validates the experimental results. However, for very high energies the simulation curves deviate from the experimental fit. This can be explained by the dynamic difference in the system expansion during melting and the relaxation during crystallization due to the difference in the confinement potential of the experiments and simulation. 

The exponents found via the MT2 and MT4 measures are consistent with the value obtained in the previous section (see Eq. \ref{eq-alphaval}). The only deviation found is for the mean exponent obtained via the $\Psi_6$ bond order method. It is significantly smaller than all the other exponents obtained (with $\langle \alpha \rangle_{\Psi 6} \simeq 0.31$ by almost $40 \% $ compared to all other exponents $\langle \alpha \rangle \simeq 0.5$) in this and the previous section. The reason for the smaller $\Psi_6$ exponent likely lies in the fact that the $\Psi_6$ bond order metric is less sensitive to lattice distortions compared to the Minkowski tensor measures and therefore less continuous in its nature, leading to a more binary, discontinuous form of defect/crystal state detection: Even for noisy data with larger distortions the Minkowski tensor method is able to distinguish more crystalline lattice structures form distorted ones, whereas the $\Psi_6$ bond order metric only finds defects and can resolve crystalline structures only for smaller distortions. This can also be seen when comparing Fig. \ref{fig-histp6} (c) with Fig. \ref{fig-betahist} (c) (respectively Fig. \ref{fig-deltahist} (c)). After the melting the $\mathrm{MT2}$ (respectively $\mathrm{MT4}$) measure starts to detect recrystallization much earlier than the $\Psi_6$ bond order metric: The histogram of the Minkowski measure shifts noticeably to the right (respectively left) whereas the bond order metric histograms start to shift to crystalline values only at much later times. 

The bond order metric is obviously more binary in nature, allowing high values only for fairly perfect crystal structure and then changing rapidly to low values for distorted crystal structure. The Minkowski tensor metrics provide a means to probe and resolve the crystal structure continuously between those extremes. This allows the Minkowski tensor measures to confirm the scaling relation Eq. \ref{eq-defectenergy} for energy levels one order of magnitude  larger (see also Fig. \ref{fig-defect_compare}) than for the $\Psi_6$ bond order metric and to confirm the FDS theory with unprecedented scrutiny.

\begin{table}
\caption{\label{tab-defect} Power law exponent $\alpha$ for the defect fraction-energy scaling in Fig. \ref{fig_defect_energy} measured via the MT2 (\ref{subsec-beta}, $\beta_{\mathrm{thresh}}=0.81$), MT4 (\ref{subsec-delta}, $\Delta_{\mathrm{thresh}}=0.18$) and $\Psi_6$ (\ref{subsec-psi6}, $\Psi_{6,\mathrm{thresh}}=0.5$) methods in Eq. \ref{eq-defectenergy} $N_T / N \propto E^{2 \alpha/\left( 1+\alpha \right)}$, for experiments I-XII and the simulation (\ref{sec-exp-sim}). For the corresponding graphs consult Fig. \ref{fig_defect_energy}. The last row is the mean value of all above with the standard deviation as uncertainty.}
\begin{ruledtabular}
\begin{tabular}{ccccccc}
 &$\alpha$ ($\Psi_6$)&$\alpha$ (MT2)&$\alpha$ (MT4)\\
\hline
\rm I & 0.186 & 0.454 & 0.413 \\
\rm II & 0.405 & 0.518 & 0.593 \\
\rm III & 0.266  & 0.323 & 0.367 \\
\rm IV & 0.295  & 0.416 & 0.473 \\
\rm V & 0.362 & 0.618 & 0.605  \\
\rm VI & 0.319  & 0.453 & 0.359 \\
\rm VII & 0.336 & 0.573 & 0.317 \\
\rm VIII & 0.405 & 0.526 & 0.418  \\
\rm IX & 0.358  & 0.563 & 0.844  \\
\rm X & 0.241  & 0.254 & 0.328 \\
\rm XI & 0.263  & 0.103 & 0.677  \\
\rm XII& 0.304  & 0.550 & 0.234  \\
\rm S & 0.303  & 0.504 & 0.471 \\
\hline
$\langle$ \rm I...XII $\rangle$ & 0.31 $\pm$0.06 & 0.52 $\pm$ 0.18 & 0.47 $\pm$0.17  \\
\end{tabular}
\end{ruledtabular}
\end{table}

\section{Discussion and Conclusion \label{sec-conclusion}}
Employing Minkowski Tensor methods to the recrystallization process of experiments and simulation of two-dimensional complex plasma systems supports the Fractal Domain Structure (FDS) phase transition theory \cite{12o} based on the kinetic theory of Frenkel \cite{13o}. The analysis of the experimental and simulation data showed a scaling behavior in crystalline self-similar domains that is not consistent with the prominent KTHNY theory of phase transitions. The results of the Minkowski tensor analysis are consistent with the theoretically predicted power laws obtained from the scale-free theory and provide higher accuracy compared to results obtained by the commonly used bond order metric $\Psi_6$ due to their capability to detect differences in defect fraction even for very high energies. Further, it is superior to the simple counting of paired 5/7-dislocations since it provides a more reliable statistic due to the much larger number of detected defects. All of the power law exponents measured via Minkowski tensor metrics are consistent for all experiments and a simulation. Furthermore, they are also consistent for two different theoretical predictors: The scale-free behaviour between defect fraction and particle energy, and the fractal relation between domain area and circumference. Summarized, this scale-free phase transition does not depend on experimental parameters but rather seems to be an inherent, universal feature of two-dimensional phase transitions as analyzed here.

The scaling relation introduced in Eq. \ref{eq-fractal} is confirmed by all experiments and the simulation for all applied defect measures ($\Psi_6$ bond order metric, MT2 and MT4 measure). The straight lines in the log-log plots (Fig. \ref{fig-area_length}) are reproduced extremely well. The power law exponents are consistent for all measures and vary only marginally with changes of parameters in the methods of identifying defects and changes in the parameters of the DBSCAN clustering algorithm applied to measure the circumference and the area of crystalline domains. The DBSCAN clustering algorithm applied in this work measures this fractal behaviour very precise and reproduces it even for the conventional $\Psi_6$ bond order method. For this method deviations from the fractal behavior were found in an earlier study \cite{knapek_rec} where domain circumference and area where calculated by counting of particles. The DBSCAN method seems to be more accurate (holes in defect lines do not play an important role) and less tedious than only counting defects.

Also, the power law Eq. \ref{eq-defectenergy} could be reproduced in with deviations only for very low and very high energies. This was already reproduced in a previous study, however only by counting paired dislocation. The extended analysis in this work validated the predicted power law more rigorously, since the applied continuous measures ($\Psi_6$ bond order metric, MT2 and MT4 measure) provide a significantly higher number of points in the defect-energy diagrams (Fig. \ref{fig_defect_energy}) for the statistical analysis. The saturation in these diagrams can be explained by fact that for high energies all lattice sites in the system are distorted to fluid levels and an upper limit is reached. For low energies we only observe small deviations due to the thermodynamic occurrence of defects that are not domain boundaries and because of particles that leave the plane of observation due to oscillation in the vertical directions \cite{vert1, vert2}. This causes artificial defects that can also be observed in the movies shown in the supplemental material \cite{supp_rec}. The power law exponents found for Eq. \ref{eq-defectenergy} are consistent with those for Eq. \ref{eq-fractal} for the Minkowski tensor methods. However, while the $\Psi_6$ measure reproduces a consistent exponent for Eq. \ref{eq-fractal} it yields a significantly smaller one for Eq. \ref{eq-defectenergy}. This is due to the fact that the $\Psi_6$ measure is more binary in its nature than the Minkowski tensor measures. Therefore the dynamic range of the measured defect fraction is smaller which is reflected in the slope and the power law exponent. The single measurement of the exponent $\alpha$ that does not fit into the other measurements in this work is however in the same range as the $\alpha$ obtained in a previous study \cite{knapek_rec} by considering defects only as 5/7-dislocations in a completely discrete fashion. The difference in these exponents might arise from this discreteness in counting defects in comparison to the more continuous Minkowski tensor methods. Here the Minkowski Tensor methods show promising potential for the analysis of crystal distortions: Where the $\Psi_6$ bond order metric only scales over one order of magnitude and fails to detect changes in crystal defect numbers for very high energies, the Minkowski tensor methods provide one more order of magnitude in scaling range. Due to its continuous nature also the joint defect lines, forming the boarders of crystalline domains, can readily be detected leading to a more precise verification of the fractal relation \ref{eq-arealength} compared to defect detection via the $\Psi_6$ bond order metric.

This study gives further evidence that Minkowski tensor methods are a powerful tool for morphological characterization of point sets. They are superior to conventional analysis methods in various respects. Minkowski tensor analysis is able to quickly reveal new aspects of interest in data, it is founded on a solid mathematical framework, however it still provides easily interpretable results.

\section*{ACKNOWLEDGMENTS \label{sec:ack}}
We thank I. Laut for carefully checking this manuscript. We thank C. Durniak for providing the simulation data. The free software Qhull (http://www.qhull.org/) was used to compute the Voronoi tessellation. Qhull uses the Quickhull algorithm for computing the convex hull. Some Minkowski Tensors are calculated based on code of the free software package Karambola (provided at http://theorie1.physik.uni-erlangen.de/research/karambola/). C. Knapek was financed by DLR/BMWi FKZ 50WM1441, A.B\"obel was funded by the StMWi.
\section*{\label{sec:citeref}}
\input{manuscript.bbl}
\end{document}

%% file: manuscript.bbl
%